\documentclass[aps,10pt,prd,groupedaddress,nofootinbib,notitlepage,eqsecnum,preprintnumbers]{revtex4-1}

\usepackage[utf8]{inputenc}
\usepackage{hyperref}
\usepackage{xcolor}
\usepackage{graphicx}
\usepackage{amsmath,amssymb,amsbsy,amstext,amsthm}
\usepackage{mathtools}
\usepackage{amsfonts}
\usepackage{comment}
\usepackage{bm}
\usepackage{tabularx}
\usepackage{array,collcell}

\newcolumntype{C}{>{\collectcell\docellC}c<{\endcollectcell}}
\newcolumntype{L}{>{\collectcell\docellL}c<{\endcollectcell}}
\newcolumntype{R}{>{\collectcell\docellR}c<{\endcollectcell}}

\makeatletter
\providecommand{\@nameedef}[1]{\expandafter\edef\csname#1\endcsname}
\newcommand{\docell}[2]{%
  \sbox\equalizedtablebox{#2}%
  \ifdim\wd\equalizedtablebox>\@nameuse{finallen\theequalizedtable}\relax
    \global\@nameedef{finallen\theequalizedtable}{\the\wd\equalizedtablebox}%
  \fi
  \makebox[\@nameuse{startinglen\theequalizedtable}][#1]{#2}%
}
\newcommand{\docellC}[1]{\docell{c}{#1}}
\newcommand{\docellL}[1]{\docell{l}{#1}}
\newcommand{\docellR}[1]{\docell{r}{#1}}
\newcounter{equalizedtable}
\newsavebox\equalizedtablebox
\newenvironment{equalizedtabular}[2][c]
  {%
   \stepcounter{equalizedtable}%
   \global\@namedef{finallen\theequalizedtable}{0pt}%
   \@ifundefined{startinglen\theequalizedtable}
    {\@namedef{startinglen\theequalizedtable}{5em}}{}
   \tabular[#1]{#2}%
  }
  {%
   \endtabular
   \begingroup\edef\x{\endgroup
     \write\@auxout{%
       \global\noexpand\noexpand\noexpand\@namedef{startinglen\theequalizedtable}%
     {\@nameuse{finallen\theequalizedtable}}%
   }}\x
  }
\makeatother

\begin{document}
\preprint{YITP-20-70, IPMU20-0053}
\title{Induced gravitational waves as a probe of thermal history of the universe}

\author{\textsc{Guillem Dom\`enech$^{a}$}}
    \email{{domenech}@{thphys.uni-heidelberg.de}}
\author{\textsc{Shi Pi$^{b,c}$} }
    \email{{shi.pi}@{ipmu.jp}} 
\author{\textsc{Misao Sasaki$^{b,d,e}$} }
    \email{{misao.sasaki}@{ipmu.jp}}

\affiliation{$^{a}$\small{Institut f\"ur Theoretische Physik, Ruprecht-Karls-Universit\"at Heidelberg, Philosophenweg 16, 69120 Heidelberg, Germany}\\
      $^{b}$\small{Kavli Institute for the Physics and Mathematics of the Universe (WPI), Chiba 277-8583, Japan}\\
      $^{c}$\small{CAS Key Laboratory of Theoretical Physics,
      Institute of Theoretical Physics,
      Chinese Academy of Sciences, Beijing 100190, China}\\
      $^{d}$\small{Center for Gravitational Physics, Yukawa Institute for Theoretical Physics, Kyoto University, Kyoto 606-8502, Japan}\\
      $^{e}$\small{Leung Center for Cosmology and Particle Astrophysics, National Taiwan University, 
      Taipei 10617, Taiwan}
    }

\begin{abstract}
The scalar perturbation induced gravitational waves are a probe of the primordial density 
perturbation spectrum on small scales. 
In this paper, we show that they can also probe the thermal history of the universe. 
We assume the universe underwent a stage with a constant equation of state parameter $w$, followed by
the radiation-dominated stage of the conventional big bang universe.
We find that the infrared slope of the power spectrum of the induced stochastic gravitational wave 
background for decelerating cosmologies is related to the equation of state of the universe. 
Furthermore, the induced gravitational wave spectrum has in general a broken power-law shape 
around the scale of reheating.
 Interestingly, below the threshold $w=0$ of the equation of state parameter,
 the broken power-law presents a peak for a Dirac delta peak in the scalar spectrum. For a finite width peak, the threshold changes to $w=-1/15$ depending on the value of the width.
 In some cases, such a broken power-law gravitational wave spectrum may
  degenerate to the spectrum from other sources like phase transitions or global cosmic strings.
\end{abstract}
 \maketitle

\section{Introduction}

The first detection of gravitational waves (GWs) from a binary black hole merger by LIGO \cite{Abbott:2016blz} opened a new door to explore cosmology. For instance, there is the possibility that the first detection of GWs came from the merger of primordial black holes (PBHs) \cite{Bird:2016dcv,Sasaki:2016jop}, which were formed by the collapse of large primordial fluctuations in the early universe (e.g. see Ref.~\cite{Sasaki:2018dmp} for a review). The observational window for gravitational wave cosmology will get wider as forthcoming ground and space based GWs detectors, such as LISA \cite{Audley:2017drz}, Taiji \cite{Guo:2018npi}, Tianqin \cite{Luo:2015ght}, DECIGO \cite{Seto:2001qf,Yagi:2011wg}, AION/MAGIS \cite{Badurina:2019hst}, ET \cite{ET} and PTA \cite{Lentati:2015qwp,Shannon:2015ect,Arzoumanian:2015liz,Qin:2018yhy}, will broaden the range of amplitudes and frequencies. For example, using cosmologists' notation, LISA and DECIGO might respectively be senstive down to $\Omega_{\rm GW}\sim 10^{-14}$ and $\Omega_{\rm GW}\sim 10^{-16}$, in the frequency range of $10^{-5}-10^{-1}\,{\rm Hz}$ and $10^{-3}-10\,{\rm Hz}$, as illustrated by the power-law integrated sensitivity curves of Refs.~\cite{Thrane:2013oya,Moore:2014lga}.

Importantly, any detection of GWs with a cosmological origin will give access to periods in our universe opaque to electromagnetic radiation, as GWs essentially propagate freely after their generation. This means that we might be able to explore the physics of the universe much before big bang nucleosynthesis. Also, from the observations of the cosmic microwave background (CMB) by Planck \cite{Akrami:2018odb} we have strong evidence that the initial conditions for the successful hot big bang cosmology were set by inflation \cite{Brout:1977ix,Starobinsky:1979ty,Guth:1980zm,Sato:1980yn}. However, little is known about the last stages of the inflationary period, what followed and how the standard radiation domination was reached. GWs may provide a way to test these unexplored regimes in the history of the universe.

Sources of cosmological GWs during these periods include phase transitions \cite{Kosowsky:1992vn,Kamionkowski:1993fg,Apreda:2001us,Grojean:2006bp,Caprini:2007xq,Caprini:2009fx,Caprini:2009yp,Hindmarsh:2013xza,Huang:2016odd,Jinno:2016vai,Chao:2017vrq,Cai:2017tmh,Cutting:2018tjt}, topological defects \cite{Vilenkin:1981zs,Vachaspati:1984gt,Krauss:1991qu,Damour:2000wa,Fenu:2009qf,Kamada:2015iga,Cui:2017ufi,Cui:2018rwi,Bettoni:2018pbl}, reheating/preheating after inflation \cite{Tashiro:2003qp,Easther:2006vd,Dufaux:2007pt,GarciaBellido:2007dg,Kuroyanagi:2015esa,Kuroyanagi:2017kfx,Liu:2017hua}, axionic resonant instabilities \cite{Kitajima:2018zco}, quantum fluctuations during inflation \cite{Guzzetti:2016mkm,DEramo:2019tit,Blasi:2020wpy}, etc. Due to the homogeneous and isotropic nature of the universe and the large number of sources, cosmological GWs will appear to the detector as a background noise or, in other words, as an isotropic stochastic gravitational wave background (SGWB) (see Ref.~\cite{Caprini:2018mtu} for a review). This GW spectrum carries information about the mechanism and time of generation and quite often presents itself as one or two broken power-laws around a characteristic scale or frequency \cite{Kuroyanagi:2018csn,Caprini:2019pxz}. For example, the spectrum of GWs generated by a first order phase transition presents two peaks at the scale corresponding to the size of the bubble (for bubble collisions) and the eddy (for sound waves) \cite{Cai:2017tmh,Kuroyanagi:2018csn}. Now, it is important to note that the detection of the SGWB often relies on the power spectrum template one is looking for in the data \cite{Kuroyanagi:2018csn,Caprini:2019pxz}. Thus, it is crucial to extensively investigate possible sources in order to classify differences and degeneracies among models.

In this regard, an important source of cosmological GWs is the so-called scalar induced SGWB \cite{tomita,Matarrese:1992rp,Matarrese:1993zf,Matarrese:1997ay,Carbone:2004iv,Ananda:2006af,Baumann:2007zm},  which has received a lot of attention recently \cite{Alabidi:2012ex,Alabidi:2013wtp,Hwang:2017oxa,Espinosa:2018eve,Kohri:2018awv,Cai:2018dig,Bartolo:2018rku,Inomata:2018epa,Yuan:2019udt,Inomata:2019zqy,Inomata:2019ivs,Chen:2019xse,Yuan:2019wwo,DeLuca:2019ufz,Tomikawa:2019tvi,Gong:2019mui,Inomata:2019yww,Yuan:2019fwv,Hwang:2017oxa,Domenech:2017ems,DeLuca:2019ufz,Gong:2019mui,Inomata:2019yww,Yuan:2019fwv,Domenech:2019quo,Ota:2020vfn,Cai:2019jah,Yuan:2019wwo,Cai:2019elf,Cai:2019amo,Bhattacharya:2019bvk}, mainly due to that fact that $(i)$ it is always generated given some primordial density fluctuations, which we know are there from the CMB, and $(ii)$ it is an essential counterpart of the PBH scenario. The detectability of the induced SGWB depends very much on the amplitude squared of the primordial density fluctuations since it comes from second order terms in cosmological perturbation theory \cite{Kodama:1985bj,Mukhanov:1990me,Noh:2004bc,Hwang:2007ni,Ananda:2006af,Baumann:2007zm}. However, note that the CMB does not provide any substantial constraint\footnote{The CMB gives a very good constraint on the shape of the primordial spectrum on the largest scales (from wavenumber $k\sim 7\cdot 10^{-4} {\rm Mpc}^{-1}$ to $k\lesssim 0.2{\rm Mpc}^{-1}$). Other constraints on smaller scales come from the current non-observation of PBHs \cite{Bugaev:2010bb,Sato-Polito:2019hws}.} on scales smaller than $k\sim 0.2{\rm Mpc}^{-1}$, which correspond to scales that left the horizon towards the last e-folds of inflation. For these reasons, the induced SGWB is a probe of the primordial density fluctuations and a way to test the last stages of inflation \cite{Assadullahi:2009jc,Bugaev:2010bb,Inomata:2018epa}.

Recently, there are claims that the induced SGWB might also be a tool to test the thermal history of the universe. On one hand, Ref.~\cite{Domenech:2019quo} extended the analytic calculations of the induced SGWB for radiation and matter dominated universes (respectively with equations of state $w=p/\rho=1/3$ and $w=0$ where $p$ and $\rho$ are the pressure and energy density) to general cosmological backgrounds with arbitrary $w>0$. There, it was shown that for an adiabatic perfect fluid the shape of the peak of the spectrum depends on the value of $w$. In similar lines, Ref.~\cite{Hajkarim:2019nbx} numerically studied the induced SGWB for $w>0$ and showed that there is a substantial impact in the GW spectrum due to the change in the effective degrees of freedom, specially around the QCD and electroweak phase transition. On the other hand, in a more general set up, Ref.~\cite{Cai:2019cdl} argued that the infrared side of the GW spectrum has a universal slope given a certain $w$. Using their estimate, it was reasoned that for some values of $w<0$ the infrared tail of the spectrum might have a red tilt. This implies that induced GWs generated in certain decelerating cosmological backgrounds might have a GW spectrum degenerate with other mechanisms. In this paper, we will investigate more carefully this claim by exploring the generation of induced GWs in cosmological backgrounds with constant deceleration, specially focusing on $w<0$. 

Cosmologies with a constant equation of state comprise the prototypical case of a perfect fluid, which could be an adiabatic perfect fluid or a self-gravitating scalar field in an exponential potential \cite{Lucchin:1984yf,Kodama:1985bj}. Scalar fields are ubiquitous in cosmology, from the field responsible for inflation \cite{Brout:1977ix,Starobinsky:1979ty,Guth:1980zm,Sato:1980yn} (including the standard model Higgs \cite{Rubio:2018ogq}) and dark energy \cite{Tsujikawa:2013fta} to axions \cite{Pajer:2013fsa} and dilatonic fields, resulting from dimensional reduction \cite{Fujii:2003pa}. In particular, exponential potentials typically appear in quintessence \cite{Copeland:1997et,Tsujikawa:2013fta} and in scale symmetric models \cite{Rubio:2017gty}. These two models, the adiabatic perfect fluid and the scalar field, give equivalent descriptions of the background expansion but differ at the level of perturbations \cite{Kodama:1985bj,Mukhanov:1990me}. On one hand, the speed of propagation of scalar perturbations, say $c_s^2$, in the adiabatic perfect fluid case is equal to the equation of state of the perfect fluid, that is $c_s^2=w$. On the other hand, the perturbations of a canonical scalar field propagate at the speed of light, i.e. $c_s^2=1$. Since we will pay particular attention to cosmological backgrounds with $w<0$, we will focus on the canonical scalar field case, as an adiabatic perfect fluid with $c_s^2<0$ is rather unphysical.

The paper is organized as follows. In section \ref{sec:powerlaw}, we review the cosmology of a canonical scalar field in an exponential potential. We also derive estimates for the infrared slope  of the induced GW spectrum assuming a peaked primordial density spectrum. In section \ref{sec:inducedgws} we provide detailed analytical calculations of the generation of induced GWs for a general value of $w>-1/3$. In section \ref{sec:radiationdomination} we compute the observed induced GW spectrum by matching our solutions during the scalar field domination to radiation domination. Lastly, section \ref{sec:conclusions} is dedicated to conclusions and further discussions on possible degeneracies with existing models of cosmological GWs. Details of the calculations can be found in the appendices.

\section{Scalar field power-law cosmology\label{sec:powerlaw}}

A convenient model for our purposes is the so-called power-law model \cite{Lucchin:1984yf}, which contains a canonical scalar field $\phi$ in an exponential potential, i.e.
\begin{align}
V(\phi)=V_0\,e^{-{\lambda\phi}/{M_{\rm pl}}}\,,
\end{align}
where $V_0$ and $\lambda$ are the free parameters of the model. Then, the total action is given by
\begin{align}
S=\int d^4x \sqrt{-g}\left\{\frac{1}{2}M_{\rm pl}^2R-\frac{1}{2}g^{\mu\nu}\partial_{\mu}\phi\partial_{\nu}\phi-V(\phi)\right\}\,,
\end{align}
where $g^{\mu\nu}$ is the metric, $M_{\rm pl}^2=1/(8\pi G)$ and $R$ is the Ricci scalar. Regarding the metric, we will consider that the universe is well described by a flat Friedman-Lema\^itre-Robertson-Walker (FLRW) metric, which in conformal time reads
\begin{align}\label{eq:FLRW}
ds^2&=a^2(\tau)\left(-d\tau^2+\delta_{ij}dx^idx^j\right)\,.
\end{align}

An exact solution to the Einstein and scalar field equations at the background level is given by
\begin{align}
a(\tau)=a_0\left(\frac{\tau}{\tau_0}\right)^{1+\beta}\quad{,}\quad {\cal H}=\frac{a'}{a}=\frac{1+\beta}{\tau}\quad{\rm and}\quad \phi'=\lambda{\cal H}M_{\rm pl}\quad{\rm where}\quad 1+\beta=\frac{2}{\lambda^2-2}\,,
\end{align}
and a prime denotes derivative with respect to conformal time, i.e. $'\equiv d/d\tau$. We have chosen this parametrization such that $\beta=0$ for a radiation dominated-like universe where $w=1/3$ (see table \ref{Tab:betaw}). This is clear by calculating the equation of state for the scalar field, which yields
\begin{align}\label{eq:beta}
w=\frac{p}{\rho}=\frac{\frac{1}{2}\phi'^2-a^2V}{\frac{1}{2}\phi'^2+a^2V}=\frac{1}{3}\frac{1-\beta}{1+\beta}\quad\Rightarrow\quad\beta=\frac{1-3w}{1+3w}\,.
\end{align}
From the scalar field equations, we also have a relation between the parameters that reads
\begin{align}\label{eq:relation}
\frac{V_0\tau_0^2a_0^{2(\beta+1)}}{M_{\rm pl}^2}=\left(2\beta+1\right)\left(\beta+1\right)\,.
\end{align}
From now on, we will consider the range $\infty> w>-1/3$ which corresponds to $-1<\beta<\infty$. It should be noted from Eq.~\eqref{eq:relation} that the range $w>1$ ($\beta<-1/2$) requires a negative potential $V_0<0$. Nevertheless, although this range of $\beta$ corresponds to unnatural potentials it is interesting to consider $w>1$ as a straightforward mathematical extension. Also, we will assume that the scalar field domination ends abruptly and the standard radiation dominated universe is reached at some (re)heating time $\tau_{\rm rh}$.

Before we get into the details of the calculations of the induced GWs in Sec.~\ref{sec:inducedgws}, it is instructive to study the rough behavior of the generation of GWs in the flat gauge, where we focus on the scalar field fluctuations. See App.~\ref{app:flatgauge} for the details. In the power-law model, the perturbations of the scalar field $\delta\phi$ behave as a massless field, just like the tensor modes $h_{ij}$. Thus, the equations of motion of the scalar field perturbations are given by the Klein Gordon equation for a massless field, namely
\begin{align}
 \delta\phi''+2{\cal H}\delta\phi'-\partial_k\partial^k\delta\phi=0\,.
\end{align}
In Fourier space, we have that $\delta\phi$ is constant on superhorizon scales ($k\ll {\cal H}$) and then oscillates and decays as $\delta\phi\propto 1/a$ on subhorizon scales ($k\gg {\cal H}$). Now, if we neglect gravitational interaction, the equations of motion for the transverse-traceless part of the metric $h_{ij}$ at second order are given by
\begin{align}\label{eq:tensormodesinstructive2}
h_{ij}''+2{\cal H}h_{ij}'-\partial_k\partial^kh_{ij}\approx M_{\rm pl}^{-2}(\partial_i\delta\phi\partial_j\delta\phi)^{TT}\,,
\end{align}
where $TT$ refers to the transverse-traceless component. As a further simplification relevant to our work, we consider that $\delta\phi$ has a  sharp peak at a certain scale $k_*$. Then the equations of motion for a given $k$-mode read
\begin{align}\label{eq:tensormodesinstructive}
h''+2{\cal H}h'+k^2h\approx \frac{k_*^2}{M_{\rm pl}^{2}}\delta\phi^2(k_*,\tau)\,.
\end{align}

First, we see that modes with $k>2k_*$ (the frequency of the source term) will not be efficiently generated, simply by momentum conservation.\footnote{Note that in the adiabatic perfect fluid case, there is a scale of narrow resonance at $k=2\sqrt{w}k_*$ as the scalar modes propagate with $c_s^2=w$. In the present case, no narrow resonance occurs.} For $k<k_*<{\cal H}$, the tensor modes have a constant source and they grow as $h\propto(k_*\tau)^2$ until the mode $k_*$ starts to oscillate after it enters the horizon at (${\cal H}(\tau_*)=k_*$). Second, we have that for $k<{\cal H}<k_*$, the source decays as $a^{-2}$ and the tensor modes evolve as $(k_*\tau)^{-2\beta}$. Note that for $\beta<0$ ($w>1/3$) the tensor modes grow on superhorizon scales. This is clear by rewriting Eq.~\eqref{eq:tensormodesinstructive} in terms of the number of e-folds, $dN={\cal H}d\tau$, which yields
\begin{align}\label{eq:efolds}
\frac{d^2h}{dN^2}+\frac{1+2\beta}{1+\beta}\frac{dh}{dN}+\frac{k^2}{{\cal H}^2}h\approx\frac{k_*^2}{{\cal H}^2M_{\rm pl}^{2}}\delta\phi^2(k_*,N)\,.
\end{align}
We see that since $\delta\phi^2\propto a^{-2}$, i.e. it decays as radiation, and the expansion rate goes as ${\cal H}^2\propto a^{-2/(1+\beta)}$, the source term for $k<{\cal H}<k_*$ evolves as $a^{-2\beta/(1+\beta)}$. Thus, for $\beta<0$ the expansion rate decays slower than the energy density of radiation and the source term for $k<{\cal H}<k_*$ grows. Therefore, we conclude that the tensor modes right before they enter the horizon at $k={\cal H}$ evolve as
\begin{align}\label{eq:superhh}
h(k,\tau)\sim {\rm constant}+(k_*\tau)^{-2\beta}\,,
\end{align}
where we already evaluated the first contribution at $k_*\tau_*\sim 1$.
Lastly, on subhorizon scales (${\cal H}<k<k_*$) we consider that the source term is negligible due to the presence of the $k^2$ term, which causes the tensor modes to oscillate and decay as $h\propto 1/a$. Matching at horizon crossing the superhorizon and subhorizon solutions during scalar domination, we roughly have that inside the horizon before the (re)heating time $\tau_{\rm rh}$ the tensor modes are given by
\begin{align}\label{eq:roughh1}
h(k_*>k>k_{\rm rh},\tau<\tau_{\rm rh})\propto \frac{e^{ik\tau}}{(k\tau)^{1+\beta}}\left({\rm constant}+\left(\frac{k_*}{k}\right)^{-2\beta}\right)\,.
\end{align}

\begin{table}
\begin{center}
\begin{equalizedtabular}{|| C || C | C | C | C | C | C | C | C ||}
    \hline
    $w$ & $-1/3$ & $-1/9$ & $-1/15$ & $0$ & $1/9$ & $1/3$ & $1$ & $\infty$
    \\ [0.5ex] 
    \hline
    $\beta$ & $\infty$ & $2$ & $3/2$ & $1$ & $1/2$ & $0$ & $-1/2$ & $-1$
    \\[0.5ex] 
    \hline
\end{equalizedtabular}
\end{center}
\caption{Relevant values of $\beta$ used in this work and their corresponding value of $w$.  \label{Tab:betaw}}
\end{table}

We can now compute the spectral tilt of the induced GW spectrum for $k_*>k>k_{\rm rh}$ using Eq.~\eqref{eq:roughh1}, where $k_{\rm rh}={\cal H}(\tau_{\rm rh})$ corresponds to the last scale that entered the horizon at (re)heating, which leads us to
\begin{align}\label{eq:GWsspectrum}
\Omega_{\rm GW}(k_*>k>k_{\rm rh},\tau<\tau_{\rm rh})=\frac{k^3}{12\pi^2{\cal H}^2}\overline{\langle|h'(k,\tau)|^2\rangle}\sim k^3(k\tau)^{-2\beta}\left({\rm constant}+\left(\frac{k}{k_*}\right)^{2\beta}\right)^2\,.
\end{align}
Note the factor $\tau^{-2\beta}$ typical of the ratio of the energy density of a radiation-like fluid and the background expansion \cite{Caprini:2018mtu}. Using the fact that modes with $k>k_{\rm rh}$ already propagate as a wave and continue to do so during radiation domination, we find that the slope of the observed GW spectrum in the power-law model for $k>k_{\rm rh}$ is given by Eq.~\eqref{eq:GWsspectrum} evaluated at $\tau_{\rm rh}$, namely
\begin{align}\label{eq:GWsspectrumslope}
\Omega_{\rm GW}(k_*>k>k_{\rm rh})\sim k^{3-2|\beta|}\,.
\end{align}
Moreover, this spectrum has to be supplemented with the GW spectrum generated on superhorizon scales before reheating, i.e. $k>k_{\rm rh}$. Thus, we can match the superhorizon solution \eqref{eq:superhh} to the linear solution during radiation domination (see App.~\ref{app:matching} for details). In this way, we obtain that
\begin{align}\label{eq:roughh22}
h(k<k_{\rm rh},\tau>\tau_{\rm rh})\propto \frac{e^{ik\tau}}{k\tau}\,.
\end{align}
Using this result, we find that for $k<k_{\rm rh}$ the observed GW spectrum slope is given by
\begin{align}\label{eq:GWsspectrumslope2}
\Omega_{\rm GW}(k<k_{\rm rh})=\frac{k^3}{12\pi^2{\cal H}^2}\overline{\langle|h'(k,\tau)|^2\rangle}\sim k^{3}\,.
\end{align}
Note that this estimate agrees with causality arguments \cite{Kamada:2015iga,Cai:2019cdl}. Furthermore, if the power spectrum of scalar fluctuations is a Dirac delta we have to further multiply by $k^{-1}$ both estimates \cite{Cai:2019cdl}.

Now, we see from Eqs.~\eqref{eq:GWsspectrumslope} and \eqref{eq:GWsspectrumslope2} that we recover the results for a radiation dominated universe, where $\Omega_{\rm GW}\sim k^3$ for a finite width peak or $\Omega_{\rm GW}\sim k^2$ for a Dirac delta peak. We also notice that for $\beta>3/2$ ($w<-1/15$) for a finite width scalar spectrum or $\beta>1$ ($w<0$) for a Dirac delta scalar spectrum, the specturm of GWs presents a peak at the scale of reheating $k_{\rm rh}$. Also we notice a degeneracy within the power-law models for the cases $|\beta|\leq 1$, that is between $w>1/3$ and $1/3>w>0$.

\section{Scalar induced gravitational waves for a peaked spectrum\label{sec:inducedgws}}

In this section, we present the detailed computations of the induced GWs in the power-law model, recovering the estimates of Sec.~\ref{sec:powerlaw}. Our starting point is a perturbed flat FLRW metric in the poisson gauge, namely
\begin{align}\label{eq:poisson}
ds^2&=a^2(\tau)\left[-(1+2\Psi)d\tau^2+\left(\delta_{ij}+2\Phi\delta_{ij}+h_{ij}\right)dx^idx^j\right]\,,
\end{align}
where $h_{ij}$ are the transverse and traceless degrees of freedom, i.e. 
\begin{align}
\delta^{ij}h_{ij}=\partial^i h_{ij}=0\,.
\end{align}

At linear order in perturbation theory we find that for a power-law scalar field dominated universe the equations of motion for $\Phi$ with a given wavenumber $k$ (see App.~\ref{app:poissongauge} for more details) are given by
\begin{align}\label{eq:epsilon}
\Phi''+2{\epsilon}{\cal H}\Phi'+k^2\Phi=0\,\quad{\rm where}\quad \epsilon=1-\frac{{\cal H}'}{{\cal H}^2}=\frac{2+\beta}{1+\beta}\,.
\end{align}
The solution that becomes constant on superhorizon scales ($k\ll {\cal H}$) and that matches with the initial conditions set by inflation reads
\begin{align}\label{eq:gravitationalpotential}
\Phi(k,\tau)=\Phi_{\rm p}(k)\,2^{\beta+3/2}\Gamma[\beta+5/2]\, (k\tau)^{-\beta-3/2} J_{\beta+3/2}(k\tau)\,,
\end{align}
where $\Phi_{\rm p}(k)$ is its primordial value which is related to the conserved curvature perturbation
on comoving slices by
\begin{align}
\Phi_{\rm p}(k)=\frac{2+\beta}{3+2\beta}{\cal R}_{{\rm p}}(k)\,.
\end{align}
Note that the gravitational potential \eqref{eq:gravitationalpotential} evolves exactly as in the case of the adiabatic perfect fluid but setting the speed of propagation to $c_s^2=1$. Thus, we should recover the results of Ref.~\cite{Espinosa:2018eve,Kohri:2018awv} for $w=1/3$ and Ref.~\cite{Domenech:2019quo} for $w>0$ once the propagation speed is set to unity.

At second order we have that the scalar modes squared\footnote{There are also scalar-tensor and tensor-tensor terms but they are subleading \cite{Gong:2019mui}.} source the linear equations of motion of the tensor modes. For a given wavenumber $k$ and polarization $\lambda$, these equations are described by
\begin{align}\label{eq:hij1}
h_{\lambda}''+2{\cal H}h_{\lambda}'+k^2h_{\lambda}=\,s_\lambda(\mathbf{k})\,,
\end{align}
where the source term is given by
\begin{align}
s_\lambda(\mathbf{k})=8\int \frac{d^3q}{(2\pi)^3}e_\lambda^{ij}(k)q_iq_j\left\{\Phi(\mathbf{q})\Phi(\mathbf{k}-\mathbf{q})+\frac{1+\beta}{2+\beta}\left[\Phi(\mathbf{q})+\frac{\Phi'(\mathbf{q})}{\cal H}\right]\left[\Phi(\mathbf{k}-\mathbf{q})+\frac{\Phi'(\mathbf{k}-\mathbf{q})}{\cal H}\right]\right\}\,,
\end{align}
and $e_{\lambda}^{ij}$ is the polarization tensor of GWs, that satisfies $\delta_{ij}e_{\lambda}^{ij}=k_ie_{\lambda}^{ij}=0$ and $e_{\lambda}^{ij}(k)e_{\lambda'}^{ij}(-k)=\delta_{\lambda\lambda'}$. Assuming that the primordial contribution to the tensor modes is negligible and using the Green's function method, one finds that the power spectrum\footnote{The power spectrum is defined by
\begin{align}
\langle {\cal R}(k){\cal R}(k')\rangle=\frac{2\pi^2}{k^3}\mathcal{P}_{{\cal R}}(k)\,\delta(\mathbf{k}+\mathbf{k}')\quad{\rm and}\quad \langle h_\lambda(k)h_\lambda(k')\rangle=\frac{2\pi^2}{k^3}\mathcal{P}_{h,\lambda}(k)\,\delta(\mathbf{k}+\mathbf{k}')\,.
\end{align}} 
of induced GWs per $\ln k$  is given by \cite{Espinosa:2018eve,Kohri:2018awv,Domenech:2019quo}
\begin{align}\label{eq:ph}
{\mathcal{P}_h(k,\tau)}=\sum_\lambda{\mathcal{P}_{h,\lambda}(k,\tau)}=8\int_0^\infty dv \int_{|1-v|}^{1+v} du&\left[\frac{4v^2-\left(1+v^2-u^2\right)^2}{4uv}\right]^2 \nonumber\\&\times\mathcal{P}_{\cal R}(kv)\mathcal{P}_{\cal R}(ku)\overline{I^2}(u,v,\beta,x)\,,
\end{align}
where an overline denotes oscillation average, we have introduced three new variables
\begin{align}\label{eq:uv}
v\equiv q/k\quad,\quad u\equiv|\mathbf{k}-\mathbf{q}|/k\quad,\quad x\equiv k\tau\,,
\end{align}
and we have defined
\begin{align}\label{eq:kernel2}
I(u,v,\beta,x)=2^{1+2\beta}\pi\frac{2+\beta}{3+2\beta}&\Gamma^2[\beta+3/2](uvx)^{-\beta-1/2}\nonumber\\&\times\left\{Y_{\beta+1/2}(x){\cal I}_J(u,v,\beta,x)-J_{\beta+1/2}(x){\cal I}_Y(u,v,\beta,x)\right\}\,,
\end{align}
where
\begin{align}\label{eq:int}
{\cal I}_{J,Y}
(u,v,\beta,x)\equiv\int_0^{x} &d\tilde x \,\tilde x^{1/2-\beta}
\left\{	
\begin{aligned}
	J_{\beta+1/2}(\tilde x)\\
	Y_{\beta+1/2}(\tilde x)
\end{aligned}
\right\}\nonumber\\&\times
\left[J_{\beta+1/2}(u\tilde x)J_{\beta+1/2}(v\tilde x)+\frac{2+\beta}{1+\beta}J_{\beta+5/2}(u\tilde x)J_{\beta+5/2}(v\tilde x)\right]\,.
\end{align}
Any analytical attempt of computing the induced SGWB essentially reduces to the calculation of \eqref{eq:int} in an analytical way. Fortunately, the integrals can be calculated in the two limiting cases of interest, on subhorizon and superhorizon scales. For scales comparable to the horizon, one requires numerical methods to study the detailed behavior of the kernel. Before we proceed, we clarify that for simplicity we refer to super/subhorizon scales those scales with respectively $k\tau\ll1$ and $k\tau\gg1$ although the terminology super/subhorizon refers to $k\ll{\cal H}$ and $k\gg{\cal H}$. The difference is most notable for $\beta\gg1$ or $\beta\sim-1$ as we have that ${\cal H}\tau=(1+\beta)$. However, the former separation, i.e. around $k\tau\sim1$, is more appropriate for approximationing the integrals as we shall see.

\subsection{Subhorizon approximation\label{subsec:subh1}}
For scales that are deep inside the horizon before reheating we have that $x\equiv k\tau\gg1$. Thus, as a good first order approximation we may take the limit of the integrals in \eqref{eq:int} to infinity. We show in App.~\ref{app:subhorizon} that the correction to this approximation always decays and it does as $x^{-1-\beta}$. Now, the definite integral of three bessel functions for $\beta>-1$ has an analytical expression first derived in Ref.~\cite{threebesselI} (also see App.~\ref{App:integralbessel}). In our case, the explicit expression for the kernel integrals on sub-horizon scales is given by
\begin{align}\label{eq:IJ}
{\cal I}_{J}
(u,v,\beta,x\gg1)=(uv)^{\beta-1/2}\frac{\left(1-y^2\right)^{\beta/2}}{\sqrt{2\pi}}\left(\mathsf{P}^{-\beta}_{\beta}(y)+\frac{2+\beta}{1+\beta}\mathsf{P}^{-\beta}_{\beta+2}(y)\right)+{\rm O}(x^{-1-\beta})
\end{align}
and
\begin{align}\label{eq:IY}
{\cal I}_Y(u,v,\beta,x\gg1)=-4(uv)^{\beta-1/2}\frac{\left(1-y^2\right)^{\beta/2}}{\left(2\pi\right)^{3/2}}\left(
	\mathsf{Q}^{-\beta}_{\beta}(y)+\frac{2+\beta}{1+\beta}\mathsf{Q}^{-\beta}_{\beta+2}(y)\right)+{\rm O}(x^{-1-\beta})\,,
\end{align}
where $\mathsf{P}^{-\beta}_\beta(y)$, $\mathsf{Q}^{-\beta}_\beta(y)$ are Legendre functions on the cut defined in the range $|y|<1$ and we have defined
\begin{align}\label{eq:y}
y\equiv1-\frac{1-(u-v)^2}{2uv}\,.
\end{align}
We present in App.~\ref{App:kernelsparticular} some explicit expressions for $\mathsf{P}^{-\beta}_\beta(y)$ and $\mathsf{Q}^{-\beta}_\beta(y)$ in terms of polynomials.

Using these results, we see that the kernel on subhorizon scales behave just like the free tensor modes in the scalar dominated universe, namely it reads
\begin{align}\label{eq:kernel33}
I(u,v,\beta,x\gg1)=x^{-(\beta+1/2)}\left(C_{1,\beta}J_{\beta+1/2}(x)+C_{2,\beta}Y_{\beta+1/2}(x)\right)\,,
\end{align}
where we have introduced for later use the coefficients
\begin{align}\label{eq:c1b}
C_{1,\beta}=-2^{1+2\beta}\pi\frac{2+\beta}{3+2\beta}&\Gamma^2[\beta+3/2](uv)^{-\beta-1/2}{\cal I}_Y(u,v,\beta,x\gg1)
\end{align}
and
\begin{align}\label{eq:c2b}
C_{2,\beta}=2^{1+2\beta}\pi\frac{2+\beta}{3+2\beta}&\Gamma^2[\beta+3/2](uv)^{-\beta-1/2}{\cal I}_J(u,v,\beta,x\gg1)\,.
\end{align}
Now, using the large argument expansion of the Bessel functions, we derive that the oscillation averaged kernel for subhorizon scales before reheating in general reads
\begin{align}\label{eq:kernelaverage}
\overline{I^2(u,v,\beta,x\gg1)}=&x^{-2(1+\beta)}\,2^{1+4\beta}\left(\frac{2+\beta}{3+2\beta}\right)^2\Gamma^4[\beta+3/2](uv)^{-2}\left(1-y^2\right)^\beta\nonumber\\&\times\left\{\left(\mathsf{P}^{-\beta}_{\beta}(y)+\frac{2+\beta}{1+\beta}\mathsf{P}^{-\beta}_{\beta+2}(y)\right)^2+\frac{4}{\pi^2}\left(
	\mathsf{Q}^{-\beta}_{\beta}(y)+\frac{2+\beta}{1+\beta}\mathsf{Q}^{-\beta}_{\beta+2}(y)\right)^2\right\}\,.
\end{align}
It is important to note that this result is not restricted to a peaked spectrum and in fact it is valid for any type of primordial spectrum. Also, the resulting averaged kernel squared \eqref{eq:kernelaverage} coincides with the results of Ref.~\cite{Espinosa:2018eve,Kohri:2018awv} for $w=1/3$ (see App.~\ref{App:kernelsparticular}) and Ref.~\cite{Domenech:2019quo} for $w>0$ once we rescaled the formulas of those references so that $c_s^2=1$. In this work, we generalize their result to negative values of the equation of state.

Before we investigate the kernel on superhorizon scales, it will be useful to investigate the infrared ($k_{\rm rh}\ll k\ll k_*$) behavior of the kernel for a peaked spectrum. In this case, we have from Eqs.~\eqref{eq:uv} and \eqref{eq:y} that $u\sim v\gg1$ and $y\sim 1$. In this limit, we find that the oscillation averaged kernel has two limiting cases as we advanced in Sec.~\ref{sec:powerlaw}. On one hand, for $-1<\beta<0$ ($w>1/3$) and $\beta\neq -1/2$ we have that
\begin{align}\label{eq:kernellimit}
\overline{I^2(v\gg1,\beta<0,x\gg1)}\approx x^{-2(1+\beta)}\,2^{1+2\beta}&\left(\frac{(2+\beta)\Gamma^2[\beta+3/2]}{\sin(\beta\pi)\Gamma[2+\beta]}\right)^2\,v^{-4-4\beta}\,.
\end{align}
The case of $\beta=-1/2$ is presented in App.~\ref{app:formulasparticularbeta} but the $v$ dependence is unchanged.
On the other hand, we see that for $\beta>0$ ($w<1/3$) the oscillation averaged kernel has a different dependence in $v$, explicitly
\begin{align}\label{eq:kernellimit2}
\overline{I^2(v\gg1,\beta>0,x\gg1)}=x^{-2(1+\beta)}\,\frac{2^{2\beta-1}}{\pi}&\left(\frac{(2+\beta)(1+\beta+\beta^2)\Gamma[\beta+3/2]}{\beta(1+\beta)^2}\right)^2\,v^{-4}\,.
\end{align}
Note that this difference in the $k$-dependence of expressions \eqref{eq:kernellimit} and \eqref{eq:kernellimit2} comes from the fact that the source term for tensor modes in Eq.~\eqref{eq:efolds} grows for $\beta<0$ and yields an additional $k^{4\beta}$ contribution to the kernel squared. This is also clear from the second term in Eq.~\eqref{eq:GWsspectrum}. 

Before ending this subsection, it should be noted that when $\beta\in \mathbb{Z}$, including $\beta=0$, the Legendre functions on the cut of the second kind present a logarithmic term that diverges in the infrared limit as $\mathsf{Q}^{-\beta}_{\beta}(y)\sim \ln(1-y)\sim \ln v^2$ (see App.~\ref{App:legendrefunctions} for the details). This will introduce a logarithmic correction to the spectral index of the GW spectrum in the infrared limit, as was pointed out in Ref.~\cite{Yuan:2019wwo} for the radiation domination case, $\beta=0$. In contrast, if $\beta\notin\mathbb{Z}$ the logarithmic correction is absent.

\subsection{Superhorizon approximation\label{subsec:superh1}}

The previous approximation is clearly not valid for scales which are superhorizon before reheating with $x=k\tau\ll1$. Nevertheless, we can perform another approximation using the fact that for a peaked spectrum we have that $u\sim v\gg1$ and $vx=k_*\tau\gg1$ close to reheating.\footnote{The correction term for $u\neq v$ can be computed by expanding around $u/v\sim1$.} To see how this approximation works, we rewrite the integral \eqref{eq:int} with a change of variables $\hat x = v \tilde x$, which leads us to
\begin{align}\label{eq:int2}
{\cal I}_{J,Y}
(v,v,\beta,x)=v^{\beta-3/2}\int_0^{vx}& d\hat x \,\hat x^{1/2-\beta}
\left\{	
\begin{aligned}
	J_{\beta+1/2}(\hat x/v)\\
	Y_{\beta+1/2}(\hat x/v)
\end{aligned}
\right\}\nonumber\\&\times
\left[J_{\beta+1/2}(\hat x)J_{\beta+1/2}(\hat x)+\frac{2+\beta}{1+\beta}J_{\beta+5/2}(\hat x)J_{\beta+5/2}(\hat x)\right]\,.
\end{align}
In this form, we see that \textit{(i)} the argument of the first Bessel function is always smaller than unity since $\hat x/v=x\ll1$ and \textit{(ii)} the limit of integration is $vx\gg1$. Thus, we can expand for small argument the first Bessel function, integrate and later expand for large $vx$. Doing so, we obtain that the integral \eqref{eq:int2} for superhorizon scales before reheating is given for general $\beta$  by
\begin{align}
{\cal I}_J(v,\beta,x\ll1,vx\gg1)\approx \frac{2^{-\beta-1/2}}{\Gamma[3/2+\beta]}\frac{3+2\beta}{1+\beta}\frac{vx}{\pi}v^{-2}
\end{align}
and
\begin{align}
{\cal I}_Y(v,\beta,x\ll1,vx\gg1)\approx& -2^{-\beta-3/2}\frac{v^{2\beta-1}}{\pi\beta(1+\beta)}\nonumber\\&\times\left(\frac{(3+2\beta)(1+\beta+\beta^2)}{(1+\beta)\Gamma[\beta+3/2]}-2^{2\beta+3}\frac{\Gamma[\beta+5/2]}{\pi(1+2\beta)}{(vx)^{-2\beta}}\right)\,.
\end{align}
The results for $\beta=-1/2$ are presented in the App.~\ref{app:formulasparticularbeta}. Also for $\beta=0$ ($w=1/3$) only the results of sec.~\ref{subsec:subh1} are needed as the universe will continue to be in a radiation dominated stage until all modes of interest are deep inside the horizon.

It is instructive for later use to expand the kernel \eqref{eq:kernel2} on superhorizon scales, which yields
\begin{align}\label{eq:kernelsuperh}
I(v,\beta,x\ll1,vx\gg1)\approx \tilde C_{1,\beta}(k)+ \tilde C_{2,\beta}(k) x^{-2\beta}
\end{align}
where
\begin{align}\label{eq:cbetas}
\tilde C_{1,\beta}(k)=\frac{(2+\beta)(1+\beta+\beta^2)}{2\beta(1+\beta)^2}v^{-2}\quad {\rm and}\quad
\tilde C_{2,\beta}(k)=-2^{1+2\beta}\frac{(2+\beta)}{\beta(1+\beta)\pi}\Gamma^2[\beta+3/2]v^{-2(1+\beta)}\,.
\end{align}
First, see how we recover the time dependence of tensor modes on superhorizon scales estimated in Eq.~\eqref{eq:superhh}. Second, in order to compute the observed GW spectrum, we need to follow these superhorizon modes after (re)heating until they are deep inside the horizon during radiation domination. In the next section we will use Eqs.~\eqref{eq:kernelaverage} and \eqref{eq:kernelsuperh} to derive the observed GW spectrum in the power-law model.

\section{Observed induced GW spectra and degeneracies\label{sec:radiationdomination}}

In Sec.~\ref{sec:inducedgws} we have computed the induced tensor modes generated during a scalar field dominated universe for a peaked spectrum. However, at some point the universe will transition to a radiation dominated universe where the standard big bang cosmology takes place. For simplicity, we will assume that the transition to radiation is sudden and the universe is reheated instantaneously. Therefore we have to match at reheating  our solutions to the linear solutions of tensor modes during radiation domination, since the source term will not be active after reheating as the scalar modes with $k_*$ have long decayed. In this way, the observed spectrum of GWs per $\ln k$ today is calculated\footnote{There is an additional factor due to change in the relativistic degrees of freedom \cite{Ando:2018qdb}. However, we will ignore as it only introduces a factor $O(1)$.} by
\begin{align}
\Omega_{\rm GW,0}h^2=\Omega_{r,0}h^2 \Omega_{\rm GW,c}\,,
\end{align}
where $\Omega_{\rm GW,c}$ is the GW spectrum evaluated at a time when the tensor modes propagate as a wave, i.e. when they are deep inside the horizon during radiation domination. Note that during radiation domination we have that $\Omega_{\rm GW,c}$ is constant since GWs behave as radiation as well. This means that we have to estimate $\Omega_{\rm GW,c}$ from Eqs.~\eqref{eq:kernelaverage} and \eqref{eq:kernelsuperh}.

We proceed with the matching as follows. First, we note that the only time dependence in the tensor modes power spectrum \eqref{eq:ph} is only through the kernel \eqref{eq:kernel2}. Thus, matching the kernels is equivalent to matching the tensor modes. Second, the kernel at the start of radiation domination, just like the tensor modes, is given by
\begin{align}\label{eq:kernelRD}
I_{RD}(x)=\frac{C_{1,RD}\sin\left(k\tau-\frac{\beta}{1+\beta}k\tau_{\rm rh}\right)+C_{2,RD}\cos\left((k\tau-\frac{\beta}{1+\beta}k\tau_{\rm rh}\right)}{k\tau-\frac{\beta}{1+\beta}k\tau_{\rm rh}}\,,
\end{align}
where we accounted for the shift in the conformal time requiring continuity of the metric at reheating and $C_{1,RD}$, $C_{2,RD}$ are obtained by matching the tensor modes and its first derivative (see App.~\ref{app:matching} for the details). At later times, after reheating and during radiation domination, that is for $k\tau\gg k\tau_{\rm rh}$, all modes of interest are deep inside the horizon and propagate as a wave. Thus, we can evaluate the spectrum of GWs as 
\begin{align}\label{eq:OMsuperh}
\Omega_{\rm GW,c}=\frac{k^2}{48{\cal H}^2}\overline{P^{RD}_h(k\tau\gg k\tau_{\rm rh})}\,,
\end{align}
where
\begin{align}\label{eq:phrd}
\overline{P^{RD}_h(k\ll k_{\rm rh},\tau\gg\tau_{\rm rh})}=\frac{8}{(k\tau)^2}\int_0^\infty dv \int_{|1-v|}^{1+v} du&\left[\frac{4v^2-\left(1+v^2-u^2\right)^2}{4uv}\right]^2 \nonumber\\&\times
\mathcal{P}_{\cal R}(kv)\mathcal{P}_{\cal R}(ku)\frac{1}{2}\left(C_{1,RD}^2+C_{2,RD}^2\right)\,,
\end{align}
and the additional factor $1/2$ is due to the oscillation average.

\begin{figure}
\centering
\includegraphics[width=0.49\textwidth]{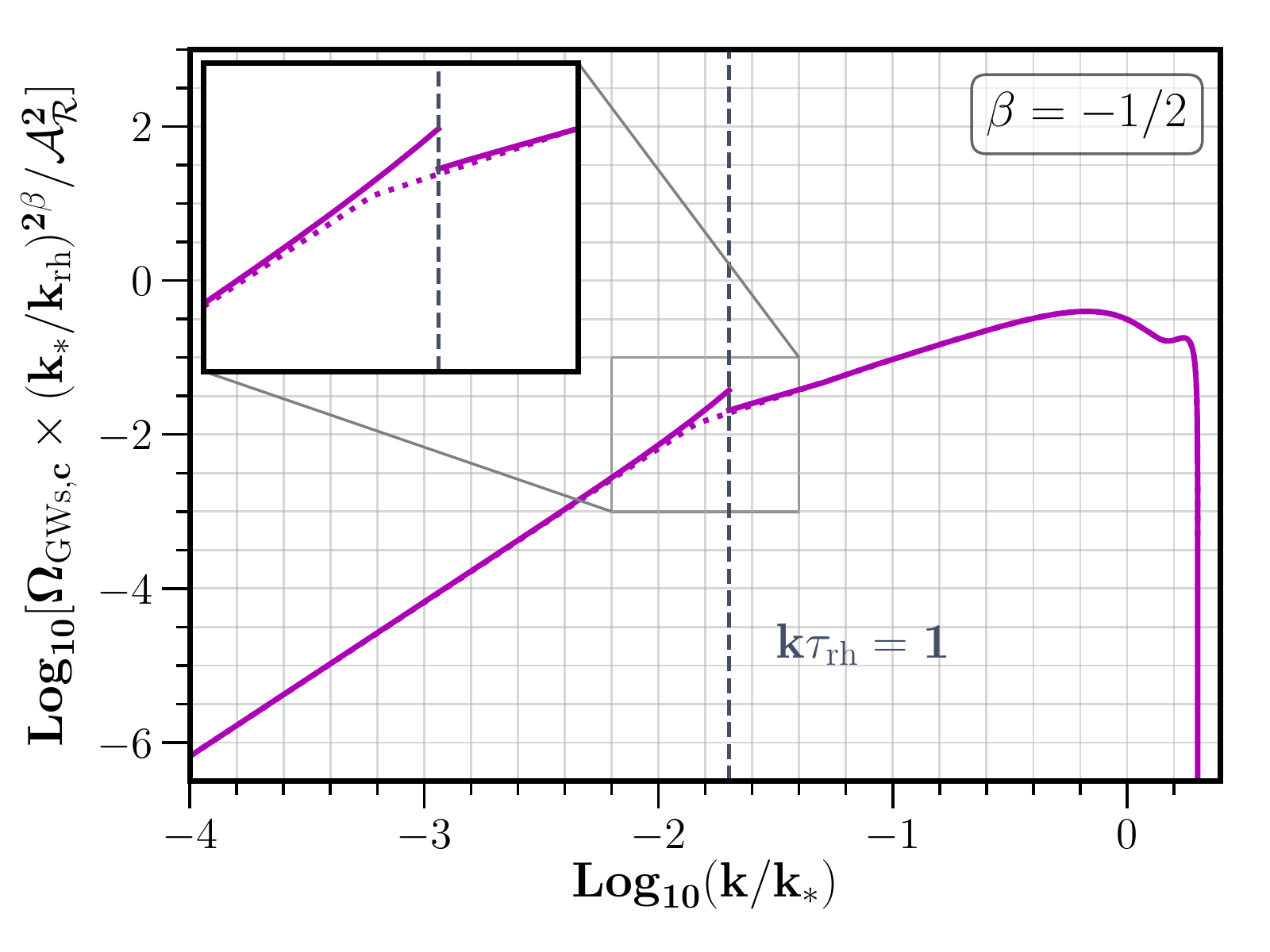}
\includegraphics[width=0.49\textwidth]{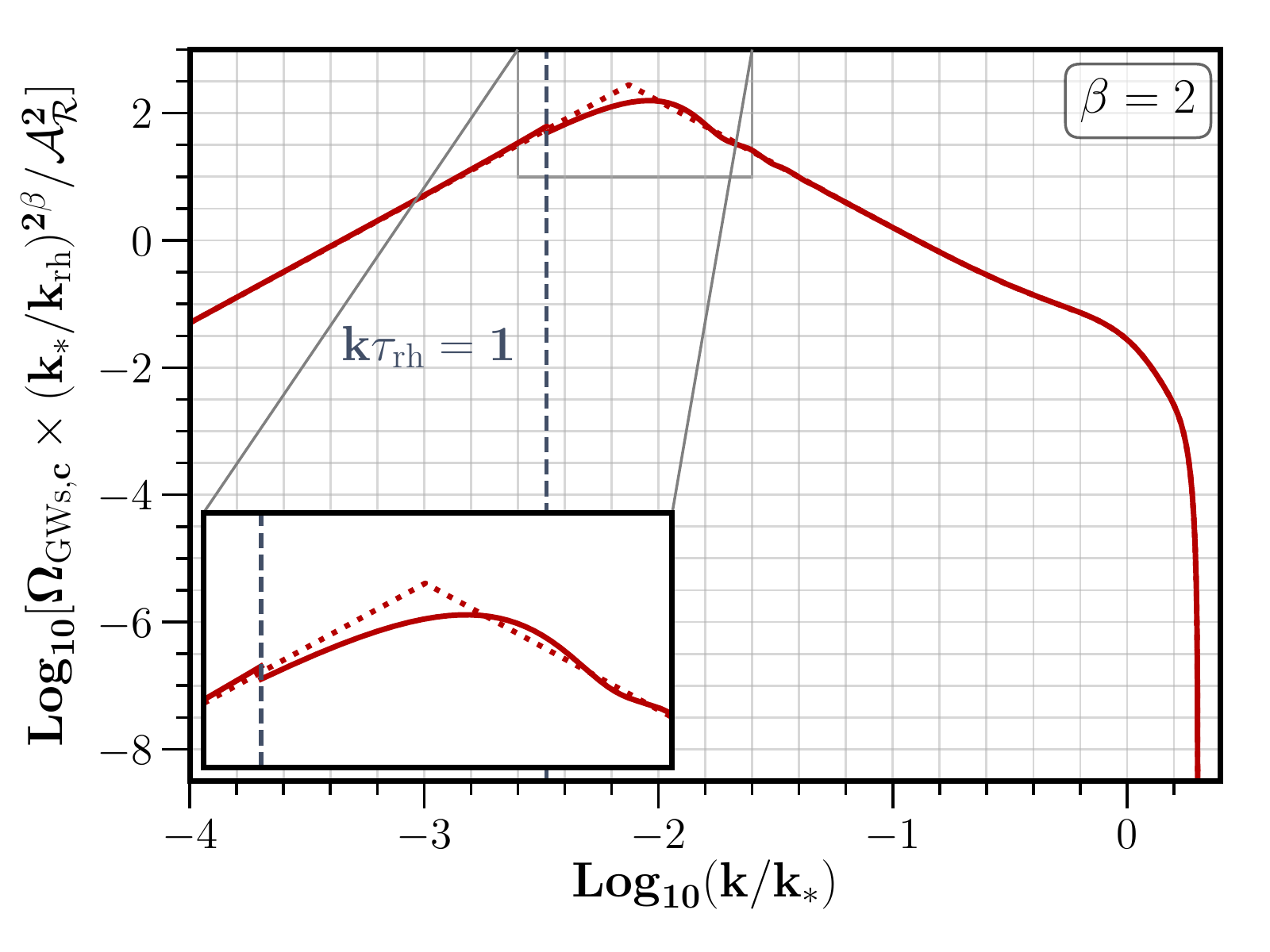}
\caption{GW spectral density for a Dirac delta power spectrum \eqref{eq:Diracdelta} for $\beta=-1/2$ (left) and $\beta=2$ (right). The solid line is the result of more accurate matching at reheating using Eqs.~\eqref{eq:kernel33}, \eqref{eq:kernelsuperh} and \eqref{eq:kernelRD} and the formulas in App.~\ref{app:matching}. The dotted line is the extrapolation from the deep subhorizon ($k\tau_{\rm rh}\gg1$ and $k\gg{\cal H}_{\rm rh}$) and far superhorizon ($k\tau_{\rm rh}\ll1$ and $k\ll{\cal H}_{\rm rh}$) regimes given by Eqs.~\eqref{eq:OMbn} and \eqref{eq:OMbp} until they cross. We have used $k_*/k_{\rm rh}=10^2$ and we have divided the spectrum by the enhancement factor $\left(k_*/k_{\rm rh}\right)^{-2\beta}$ for easier comparison between spectra. The reheating scale is the scale that last crossed the horizon at $\tau_{\rm rh}$, \textit{i.e.} $k_{\rm rh}\tau_{\rm rh}=1+\beta$. In the left figure, see how the point where the extrapolations meet is close to $k_{\rm rh}/(1+\beta)$. This could be an indication that the knee of the spectrum is near $k_{\rm rh}/(1+\beta)$ although numerical calculations are needed to confirm this. It should be noted that for $\beta<0$ the super/subhorizon approximation Eqs.~\eqref{eq:kernel33} and \eqref{eq:kernelsuperh}, respectively with $k\tau\ll1$ and $k\tau\gg1$, breaks down faster the negative the beta. This is why there is more uncertainty near $k\sim k_{\rm rh}/(1+\beta)$ when we used the full matching at reheating. In the right figure, note how the spectrum presents a peak near $k\sim k_{\rm rh}$ both in the more accurate matching and in the extrapolation. For $\beta>0$ our super/subhorizon approximation Eqs.~\eqref{eq:kernel33} and \eqref{eq:kernelsuperh} yield better results in contrast to $\beta<0$ and this explains why both approximation join well at $k\sim k_{\rm rh}/(1+\beta)$. We can also see small oscillations near the reheating scale. Numerical calculations are needed in order to see the actual behavior of the spectrum around $k\sim k_{\rm rh}/(1+\beta)$ and $k\sim k_{\rm rh}$. \label{fig:deltasmatching}}
\end{figure}

Now, on one hand, matching modes which entered the horizon much after reheating, that is modes with $k\ll k_{\rm rh}$, we have that
\begin{align}\label{eq:farsuperh}
C_{1,RD}=\tilde C_{1,\beta}+\frac{1-\beta}{1+\beta}\tilde C_{2,\beta}(k\tau_{\rm rh})^{-2\beta}\quad{,}\quad
C_{2,RD}=\frac{2\beta}{\left(1+\beta\right)^2} \tilde C_{2,\beta}(k\tau_{\rm rh})^{1-2\beta}\,,
\end{align}
where $C_{1,\beta}$ and $C_{2,\beta}$ are given in terms of $v$ and $\beta$ in Eq.~\eqref{eq:cbetas}. Recall that this superhorizon approximation is valid for $k\tau_{\rm rh}\ll 1$ and $k\ll {\cal H}_{\rm rh}$. However, for $\beta<0$ we have that ${\cal H}_{\rm rh}<1/\tau_{\rm rh}$ and, therefore, in the region with $k_{\rm rh}/(1+\beta)>k>k_{\rm rh}$ where we have to use a more accurate matching presented in App.~\ref{app:matching}. Nevertheless, see in Fig.~\ref{fig:deltasmatching} how the extrapolation of \eqref{eq:farsuperh} gives a good estimate.  

On the other hand, we have that modes which entered the horizon much before reheating, that is modes with $k\gg k_{\rm rh}$, are already oscillating. After matching we find that
\begin{align}\label{eq:deepsubh}
\frac{1}{2}\left(C_{1,RD}^2+C_{2,RD}^2\right)=\frac{1}{2}\left(C_{1,\beta}^2+C_{2,\beta}^2\right)=\left(\frac{k\tau_{\rm rh}}{1+\beta}\right)^2\overline{I^2(u,v,\beta,x_{\rm rh})}\,,
\end{align}
where we used Eqs.~\eqref{eq:c1b} and \eqref{eq:c2b}, $\overline{I^2}$ is the average subhorizon kernel squared given in Eq.~\eqref{eq:kernelaverage}. This result was expected since the subhorizon modes were already propagating as a wave just before reheating and continue to do so during radiation domination. This is also why the latter result coincides with Eqs.~\eqref{eq:ph} and \eqref{eq:kernelaverage} evaluated at reheating, namely
\begin{align}\label{eq:subh}
\Omega_{\rm GW,c}(k\gg k_{\rm rh})=\frac{k^2}{48{\cal H}^2}\overline{\mathcal{P}_h(k\gg k_{\rm rh},\tau)}\bigg|_{\tau=\tau_{\rm rh}}\quad{\rm where}\quad \tau_{\rm rh}=(1+\beta)/k_{\rm rh}\,.
\end{align}
It should be noted that this approximation is valid when $k\tau_{\rm rh}\gg 1$ and $k\gg {\cal H}_{\rm rh}$. This means that for $\beta>0$ we have that ${\cal H}_{\rm rh}>1/\tau_{\rm rh}$ and, thus, in the region with $k_{\rm rh}/(1+\beta)<k<k_{\rm rh}$ we have to use again a more accurate matching presented in App.~\ref{app:matching}. However, as shown in Fig.~\ref{fig:deltasmatching} the extrapolation of \eqref{eq:farsuperh} gives a good estimate.

\begin{figure}
\centering
\includegraphics[width=0.75\textwidth]{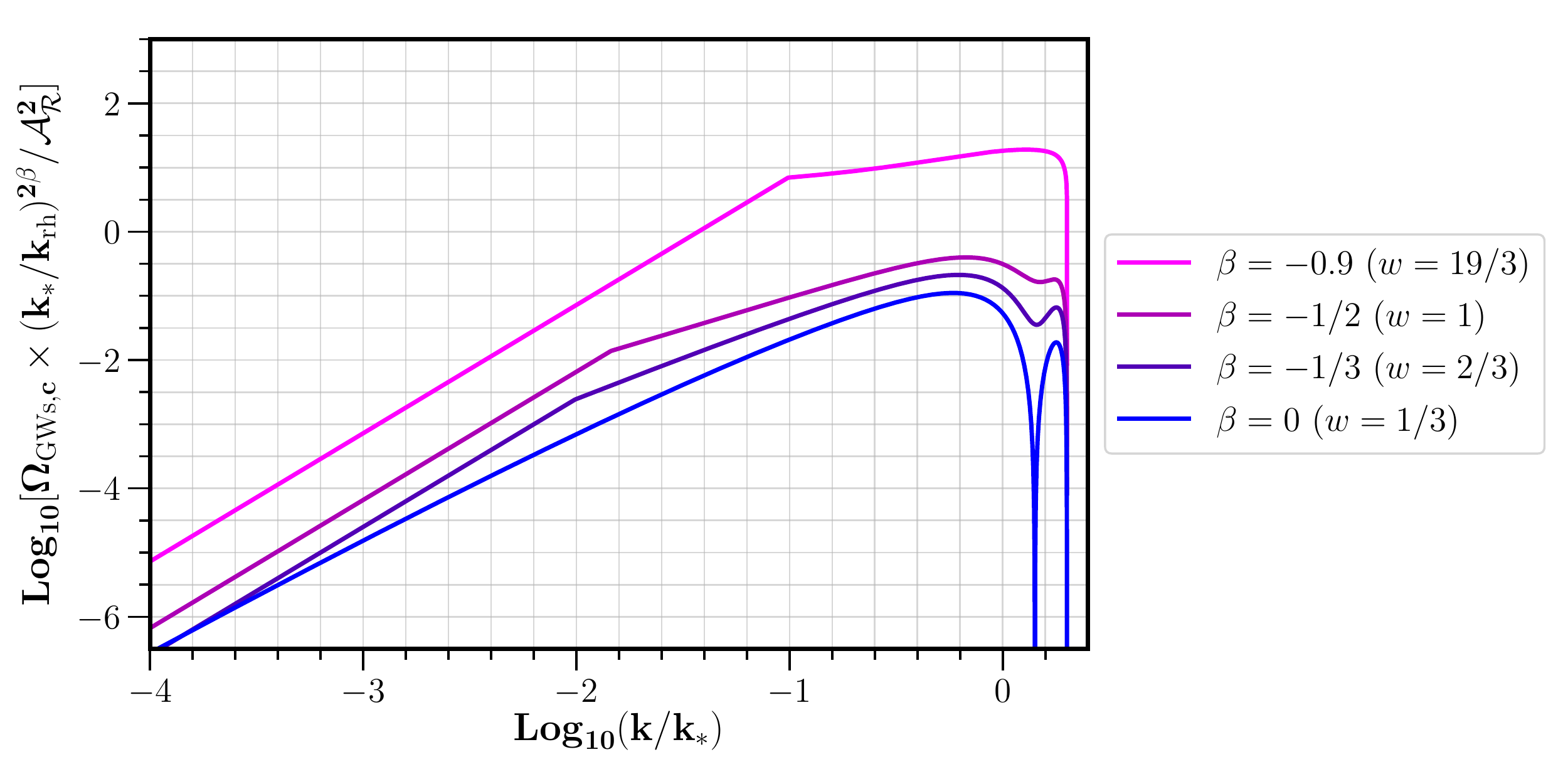}
\caption{GW spectral density for a Dirac delta power spectrum \eqref{eq:Diracdelta} and $-1<\beta\leq 0$ ($w\geq 1/3$). We have used $k_*/k_{\rm rh}=10^2$ and we have divided the spectrum by the enhancement factor $\left(k_*/k_{\rm rh}\right)^{-2\beta}$ for easier comparison between spectra. The reheating scale is the scale that last crossed the horizon at $\tau_{\rm rh}$, \textit{i.e.} $k_{\rm rh}\tau_{\rm rh}=1+\beta$. The two approximations in Eq.~\eqref{eq:OMbn} meet at $k\sim k_{\rm rh}/(1+\beta)$. However, numerical calculations are needed in order to see the actual behavior of the spectrum near $k\sim k_{\rm rh}$ and $k\sim k_{\rm rh}/(1+\beta)$. \label{fig:deltasbn}}
\end{figure}

At this point a note is in order. We expect a small correction due to the sudden reheating approximation. First, notice that from Eq.~\eqref{eq:epsilon} we have that $1<\epsilon<3$ for $-1/3<w<1$, with $\epsilon=2$ for radiation. Thus, the difference in $\epsilon$ from scalar to radiation domination is at most $1$. Furthermore, since the potential $\Phi$ oscillates rapidly inside the horizon regardless of the equation of state, we can use the WKB approximation. In this case, the correction to the amplitude is of the order of $\left(\epsilon-2\right){\cal H}_{\rm rh}/k$ relative to the dominant term, which is in general small except for modes close to $k_{\rm rh}$. This is in contrast with Refs.~\cite{Inomata:2019zqy,Inomata:2019ivs}, where they find an enhancement of the GWs spectrum due to a sudden transition from matter to radiation domination for an adiabatic perfect fluid. In their case, the gravitational potential $\Phi$ does not decay even after the horizon re-entry, and the time derivative of the gravitational potential $\Phi$ experiences an abrupt change, from zero to an oscillating function, which is the source of the enhancement.

In summary, we have derived the observed GW spectrum induced by a peaked density power spectrum during a scalar field dominated universe with $\beta>-1$ (or $w>-1/3$) and it is given by Eqs.~\eqref{eq:OMsuperh} and \eqref{eq:phrd}, complemented by Eqs.~\eqref{eq:farsuperh} and \eqref{eq:deepsubh}. It should be noted that for scales close to the reheating scale $k\sim k_{\rm rh}$, we expect that there would be small oscillations on top of our result \cite{Assadullahi:2009nf}, reflecting the first few oscillations of the tensor modes that entered the horizon just before reheating. This is clear for $\beta>0$ in the right plot of Fig.~\ref{fig:deltasmatching}. However, for $\beta<0$ one needs a more accurate approximation to the superhorizon regime since for $\beta<0$ the approximation breaks down faster near $k\tau_{\rm rh}\sim 1$ than for $\beta>0$. Nevertheless, as we have argued our estimate should give the right order of magnitude even for $k\sim k_{\rm rh}$. We will proceed to study the IR limits for a Dirac delta and a finite width primordial scalar power spectrum.

\subsection{IR limit of GW spectrum for a Dirac delta scalar spectrum\label{subsec:delta}}

\begin{figure}
\centering
\includegraphics[width=0.75\textwidth]{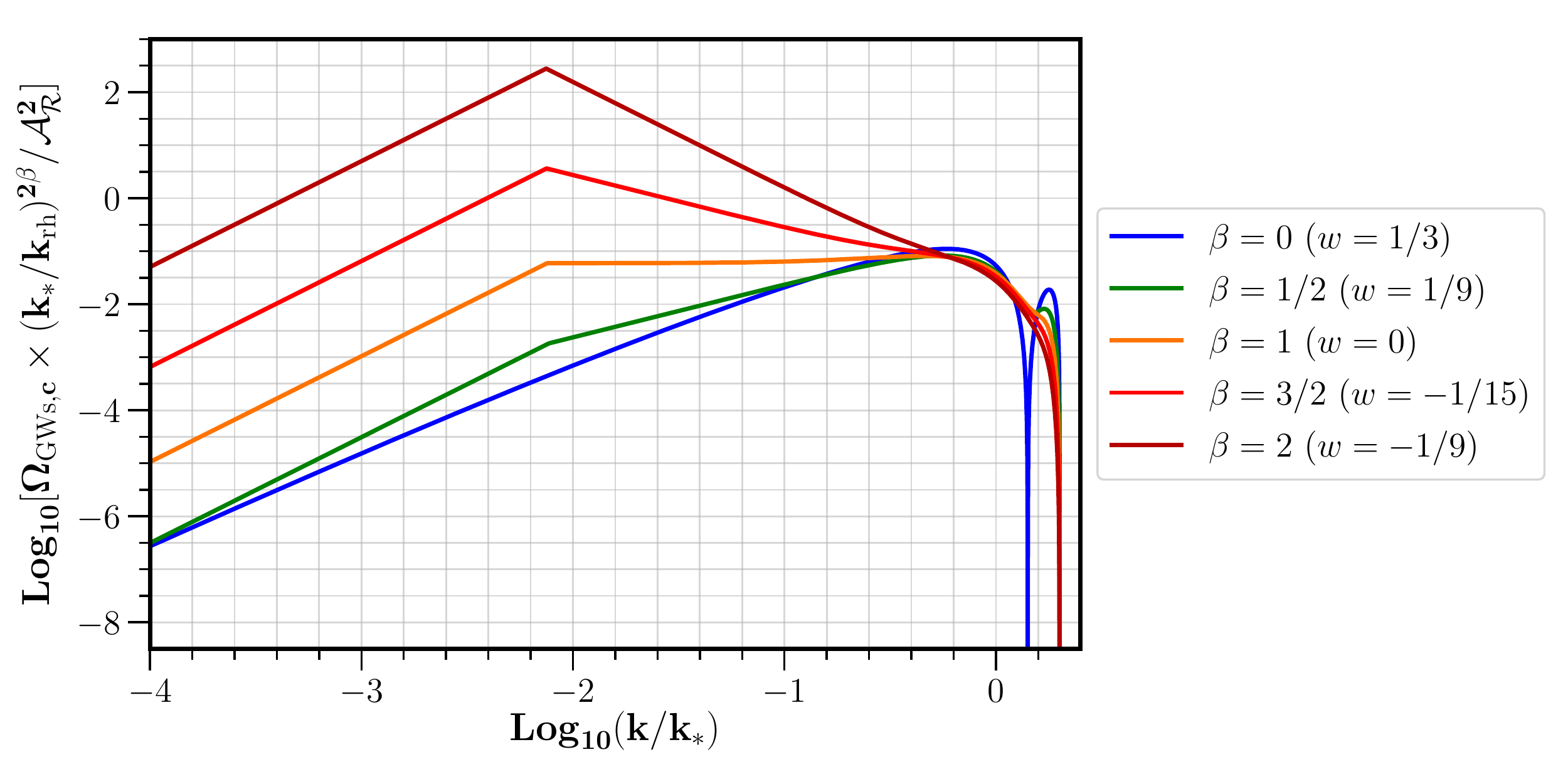}
\caption{GW spectral density for a Dirac delta power spectrum \eqref{eq:Diracdelta} and $\beta\geq 0$ ($w\leq 1/3$). We have used $k_*/k_{\rm rh}=10^2$ and we have divided the spectrum by the enhancement factor $\left(k_*/k_{\rm rh}\right)^{-2\beta}$ for easier comparison between spectra. The reheating scale is the scale that last crossed the horizon at $\tau_{\rm rh}$, \textit{i.e.} $k_{\rm rh}\tau_{\rm rh}=1+\beta$. The two approximation given in Eq.~\eqref{eq:OMbp} meet at $k\sim k_{\rm rh}$ in agreement with the full matching of App.~\ref{app:matching}. See also Fig.~\ref{fig:deltasmatching}. \label{fig:deltasbp}}
\end{figure}

Let us consider that the primordial spectrum of scalar fluctuations has a infinitely sharp peak at a scale $k_*$. Explicitly we assume that it is given by
\begin{align}\label{eq:Diracdelta}
\mathcal{P}_{\cal R}(k)={\cal A}_{\cal R}k_*\delta(k-k_*)\,.
\end{align}
In this case, we can directly use Eqs.~\eqref{eq:OMsuperh} and \eqref{eq:phrd} and evaluate the integrand at $u=v=k/k_*$. The resulting GW spectra for $-1<\beta\leq 0$ ($w\geq 1/3$) and $\beta\geq 0$ ($w\leq 1/3$) are respectively shown in Figs.~\ref{fig:deltasbn} and \ref{fig:deltasbp}. See how for $\beta>1$ there is a peak around the scale of reheating $k_{\rm rh}$. Also, note how the greater the $\beta$ the steeper the slope for $k>k_{\rm rh}$.

To have an idea of the slopes of the spectrum in the infrared regime for a Dirac delta we expand Eqs.~\eqref{eq:OMsuperh} and \eqref{eq:phrd} for $k\ll k_*$. We respectively find for $\beta<0$
\begin{align}\label{eq:OMbn}
\Omega_{\rm GW,c}(\beta<0,k\ll k_*)=\frac{{\cal A}^2_{\cal R}}{12\pi^2}&\left(\frac{2^{1+\beta}(2+\beta)\Gamma^2[3/2+\beta]}{\left(1+\beta\right)^{1+\beta}}\right)^2\left(\frac{k_{\rm rh}}{k_*}\right)^{2\beta}\nonumber\\&\times
\left\{
\begin{aligned}
&\left(\frac{2^{\beta}(1-\beta)}{\beta\left(1+\beta\right)^{1+\beta}}\right)^2\left(\frac{k_{\rm rh}}{k_*}\right)^{2\beta}\left(\frac{k}{k_*}\right)^{2} &(k\lesssim \frac{k_{\rm rh}}{1+\beta})\\
&\left(\frac{\pi}{\sin(\beta\pi)\Gamma[2+\beta]}\right)^2\left(\frac{k}{k_*}\right)^{2+2\beta}  &(k\gtrsim \frac{k_{\rm rh}}{1+\beta})
\end{aligned}
\right.\,,
\end{align}
and for $\beta>0$
\begin{align}\label{eq:OMbp}
\Omega_{\rm GW,c}(\beta>0,k\ll k_*)=\frac{{\cal A}^2_{\cal R}}{{48\pi}}&\left(\frac{(2+\beta)(1+\beta+\beta^2)}{\beta\left(1+\beta\right)^{2}}\right)^2\nonumber\\&\times
\left\{
\begin{aligned}
&\pi\left(\frac{k}{k_*}\right)^{2}\quad &(k\lesssim k_{\rm rh})\\
&\left(\frac{{2^{1+\beta}}\Gamma[\beta+3/2]}{\left(1+\beta\right)^{1+\beta}}\right)^2\,\left(\frac{k_{\rm rh}}{k_*}\right)^{2\beta}\left(\frac{k}{k_*}\right)^{2-2\beta}\quad &(k\gtrsim k_{\rm rh})
\end{aligned}
\right.\,.
\end{align}
We choose the position of the matching as the point where the two approximation cross, which is of the order of $k\sim k_{\rm rh}/(1+\beta)$ for $\beta<0$ and $k\sim k_{\rm rh}$ for $\beta>0$ and coincides with the results of the a more accurate matching at reheating derived in App.~\ref{app:matching}. For an illustration see Fig.~\ref{fig:deltasmatching}. Also note how we recover the predictions for the slopes \eqref{eq:GWsspectrumslope} and \eqref{eq:GWsspectrumslope2} for a Dirac delta scalar spectrum. Therefore, we have shown that the infrared tail of the induced GW spectrum is a probe of the thermal history of the universe.

Let us look closer at the results \eqref{eq:OMbn} and \eqref{eq:OMbp}. On one hand we find that for $\beta<0$ the induced GW spectrum has an enhancement factor $\left({k_{\rm rh}}/{k_*}\right)^{2\beta}$ due to the relative background expansion and the superhorizon growth of tensor modes. This is also the reason why the slope of the infrared tail of the spectrum for $k>k_{\rm rh}$ goes as $k^{2+2\beta}$ instead of $k^{2-2\beta}$. Furthermore, we find that the spectrum for $\beta\sim -1$ gets enhanced by an additional factor $(1+\beta)^{-2}$. See Fig.~\ref{fig:deltasbn} for an illustration with three cases, $\beta=\{-1/3,-1/2,-9/10\}$, compared to radiation domination, $\beta=0$. We conclude that for $\beta<0$ the peak of the spectrum is close to the characteristic scale of the peak in the scalar spectrum $k\sim k_*$.

On the other hand, for $\beta>0$ we see that the induced GW spectrum has a suppression factor only due to the relative background expansion. Furthermore, the spectrum for $\beta>1$ presents a peak at around the reheating scale $k_{\rm rh}$. As this case might be degenerate with other mechanisms, we can estimate the peak of the spectrum for $\beta>1$ to have an amplitude proportional to
\begin{align}
\Omega^{\rm peak}_{\rm GWs,c}&(\beta>1,k_*\gg k\sim k_{\rm rh})\approx\frac{{\cal A}^2_{\cal R}}{{48}}\left(\frac{k_{\rm rh}}{k_*}\right)^{2}\,.
\end{align}
We see that the amplitude of the peak is suppressed by a factor $\left({k_{\rm rh}}/{k_*}\right)^{2}$ independent of $\beta$. This means that for $\beta>0$ the longer the scalar field dominated stage, the smaller the amplitude of the GW spectrum. See Fig.~\ref{fig:deltasbp} for an illustration with four cases, $\beta=\{1/2,1,3/2,2\}$, compared to radiation domination, $\beta=0$. It should be noted that the GW spectrum for $w=0$ in the power-law scalar dominated universe is very different from that of a pressureless adiabatic perfect fluid as in Refs.~\cite{Assadullahi:2009nf,Inomata:2019ivs}. In the present case, GWs are generated due to the time dependence of the scalar field fluctuations (or the gravitational potentials). In the case of a pressureless adiabatic perfect fluid, where the gravitational potential is constant, GWs are created at the transition to radiation domination \cite{Inomata:2019ivs}.

Now, if we compare Eqs.~\eqref{eq:OMbn} and \eqref{eq:OMbp}, we find that the tilt of the infrared slope of the spectrum is degenerate for $|\beta|<1$. However, they seem to differ in the position of the knee which is $k\sim {k_{\rm rh}}/{(1+\beta)}$ for $\beta<0$ and $k\sim k_{\rm rh}$ for $\beta>0$, although a numerical calculation is needed to confirm the behavior for $\beta<0$. Also, if we look at left plot of Fig.~\ref{fig:deltasmatching} we observe that our approximation for $\beta<0$ is less accurate near $k\sim k_{\rm rh}/(1+\beta)$ than $\beta>0$. Thus, we expect that the detailed spectrum in this range would break the degeneracy for $|\beta|<1$. The reason is that the lower the $\beta$ the less the oscillations of the tensor modes are damped compared to the background expansion. Thus, the imprint of the oscillations for the scales close to $k_{\rm rh}$ in the spectrum will be larger for $\beta<0$. Numerical calculations are left for future work.

\begin{figure}
\centering
\includegraphics[width=0.49\textwidth]{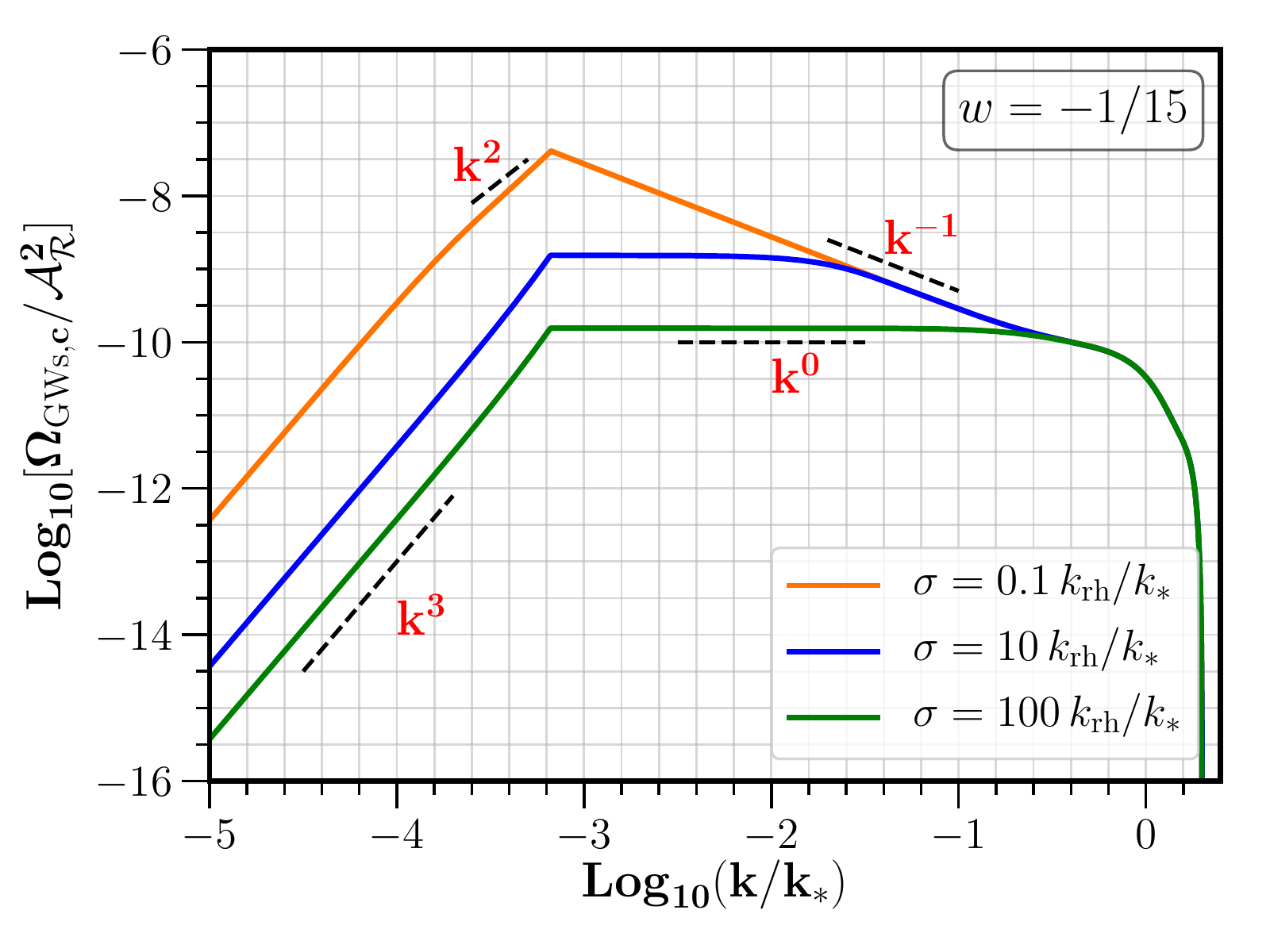}
\includegraphics[width=0.49\textwidth]{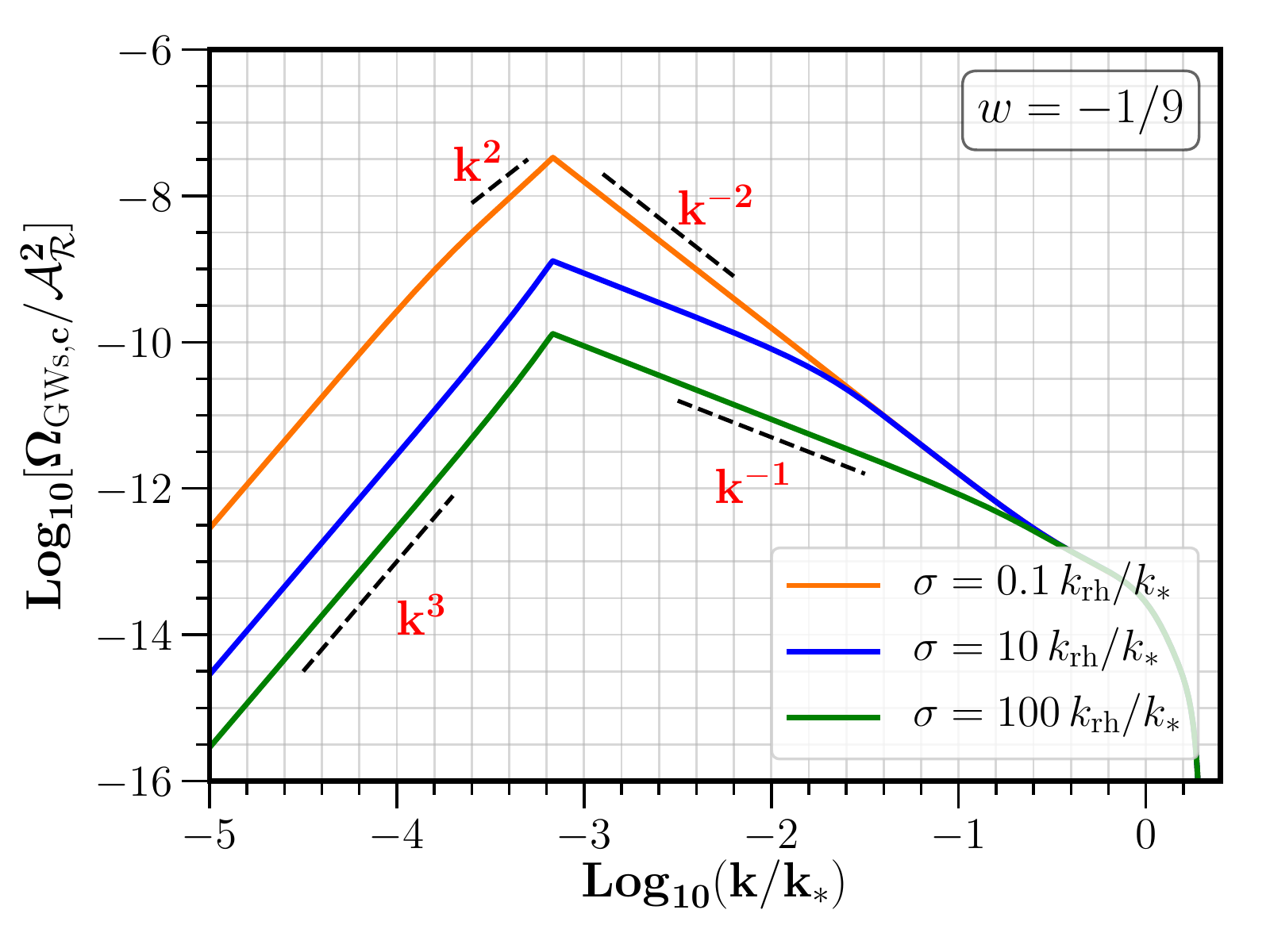}
\caption{GW spectral density for a finite width peaked power spectrum \ref{eq:lognormalpeak} where we have used $k_*/k_{\rm rh}=10^3$ and $k_{\rm rh}$ corresponds to the last scale that crossed the horizon at $\tau_{\rm rh}$. We have chosen the position of the peak as the position of the crossing between the two approximations Eqs.~\eqref{eq:subh} and \eqref{eq:OMsuperh}. We respectively plot $\beta=3/2$ ($w=-1/15$) and $\beta=2$ ($w=-1/9$) on the left and right figures and we show the spectra for width $\sigma=0.1 k_{\rm rh}/k_*,10 k_{\rm rh}/k_*,100 k_{\rm rh}/k_*$ respectively in orange, blue and green. Interestingly, the spectrum for $\beta=2$ and $\sigma=100 k_{\rm rh}/k_*$ (right figure green line) is almost degenerate with the spectrum from first order phase transitions \cite{Kuroyanagi:2018csn}. Lastly, note how the GW spectrum's $k$-dependence is increased by one additional power of $k$ when $k<\sigma\,k_*$. \label{fig:finitewidth}}
\end{figure}

\subsection{IR limit of GW spectrum for a scalar spectrum with finite width\label{subsec:finitewidth}}

In a realistic scenario the peak of the scalar spectrum has a finite width. As it was observed in Ref.~\cite{Cai:2019cdl}, such finite width of the peak affects the infrared scaling of the induced GW spectrum. For instance, modes that entered the Hubble horizon during radiation domination present an infrared scaling of the GW spectral density proportional to $k^2$ for a Dirac delta and $k^3$ for a broad peak. Furthermore, the finiteness of the width introduces a new scale at which the infrared scale may change. Interestingly, if the dimensionless width of the peak, say $\sigma$, is smaller than $1$, the induced GW spectrum will transition from a $k^3$ to a $k^2$ infrared scaling at around $k/k_*\sim\sigma$. We expect similar conclusions for a general equation of state. However, as we will see, the presence of the reheating scale $k_{\rm rh}$ introduces richer structure in the GW spectrum.

We assume that the finite width peak of the scalar spectrum is parameterized by a log-normal distribution, that is
\begin{align}\label{eq:lognormalpeak}
\mathcal{P}_{\cal R}(k)=\frac{\mathcal{A_{\cal R}}}{(2\pi)^{3/2}2\sigma k^3}\exp\left[-\frac{\ln^2(k/k_*)}{2\sigma^2}\right],
\end{align}
where $\sigma$ is the dimensionless width of the peak. GWs induced by a peaked scalar spectrum such as Eq.~\eqref{eq:lognormalpeak} are studied in detail in Ref.~\cite{Pi:2020xxx}. Here we directly use their result for $\sigma\ll 1$ which reads
\begin{align}\label{eq:omegafinitewidth2}
\Omega_{\text{GW},\sigma}(k)=\text{erf}\left(\frac{1}{\sigma}\sinh^{-1}\frac{k}{2k_*}\right)
\Omega_{\text{GW},\delta}(k),
\end{align}
where $
{\rm erf}(x)$ is the error function and $\Omega_{\text{GW},\delta}(k)$ is the GW spectrum induced by a $\delta$-function peak given by Eqs.~\eqref{eq:OMsuperh} and \eqref{eq:subh}. For a broad peak ($\sigma\gtrsim1$), the near-peak shape of the induced GWs is also log-normal, with a width of nearly $\sigma/\sqrt{2}$, which is a reflection of the secondary nature of the induced GWs. We refer the reader to Ref.~\cite{Pi:2020xxx} for further details. 

Now, let us focus in the sharp peak and infrared limits, i.e. $\sigma\ll 1$ and $k/k_*\ll 1$, where expression \eqref{eq:omegafinitewidth2} reduces to
\begin{align}\label{eq:omegafinitewidth}
\Omega_{\text{GW},\sigma\ll1}(k\ll k_*)\approx\text{erf}\left[\frac{k}{2k_*\sigma}\right]
\Omega_{\text{GW},\delta}(k\ll k_*)\,.
\end{align}
From Eq.~\eqref{eq:omegafinitewidth} we see the role of the new scale $\sigma$. When $\sigma\ll k/k_*$ we have that $\text{erf}\left[{k}\big/({2k_*\sigma})\right]\sim 1$ and we recover the results for the Dirac delta of Sec.~\ref{subsec:delta}. In contrast, for $\sigma\gg k/k_*$ we find that $\text{erf}\left[{k}\big/({2k_*\sigma})\right]\sim {k}\big/({2k_*\sigma})$ and the $k$-dependence of the spectral density changes by one additional power of $k$. In addition to that, we have two different possibilities since the GW spectrum from a Dirac delta Eqs.~\eqref{eq:OMbn} and \eqref{eq:OMbp} presents a different infrared scaling for $k>k_{\rm rh}$ and $k<k_{\rm rh}$. First, when $\sigma>k_{\rm rh}/k_*$ the change in the GW spectrum's slope occurs for $k>k_{\rm rh}$. This case is particularly interesting since the transition is from $k^{3-2|\beta|}$ to $k^{2-2|\beta|}$ and depends on the equation of state $\beta$. Second, for $\sigma<k_{\rm rh}/k_*$ the change happens for $k<k_{\rm rh}$ and we recover the results of Ref.~\cite{Cai:2019cdl} during radiation domination, namely from $k^3$ to $k^2$ independent of $\beta$. Thus, we conclude that for $\sigma\ll1$ the infrared tail of the induced GW spectrum goes as
\begin{align}\label{eq:summaryfinitewidth}
\Omega_{\rm GWs,\sigma}(k\ll k_*)\sim\left\{
\begin{aligned}
&k^{3} & (\sigma\,k_*>k_{\rm rh}>k)\\
&k^{3-2|\beta|} &(\sigma\,k_*>k>k_{\rm rh})\\
&k^{2-2|\beta|}& (k>\sigma\,k_*>k_{\rm rh})
\end{aligned}
\right.
\end{align}
or
\begin{align}
\Omega_{\rm GWs,\sigma}(k\ll k_*)\sim\left\{
\begin{aligned}
&k^{3} & (k_{\rm rh}>\sigma\,k_*>k)\\
&k^{2} & (k_{\rm rh}>k>\sigma\,k_*)\\
&k^{2-2|\beta|} &(k>k_{\rm rh}>\sigma\,k_*)
\end{aligned}
\right.\,,
\end{align}
respectively if $\sigma>k_{\rm rh}/k_*$ or $\sigma<k_{\rm rh}/k_*$.
These two possibilities are illustrated in Fig.~\ref{fig:finitewidth} for $\beta=3/2$ ($w=-1/15$) and $\beta=2$ ($w=-1/9$) with three different widths, concretely $\sigma=0.1\,k_{\rm rh}/k_*$, $10\, k_{\rm rh}/k_*$ and $100 \,k_{\rm rh}/k_*$. See how depending on the value of $\sigma$ there is a knee in the power spectrum where the slope changes by $1$, in addition to the knee at the scale of reheating. 

\section{Discussion and conclusions \label{sec:conclusions}}

The possible detection of stochastic gravitational wave backgrounds with cosmic origin by future space based detectors, e.g. LISA \cite{Audley:2017drz}, Taiji \cite{Guo:2018npi}, Tianqin \cite{Luo:2015ght}, DECIGO \cite{Seto:2001qf,Yagi:2011wg}, AION/MAGIS \cite{Badurina:2019hst}, ET \cite{ET} or PTA \cite{Lentati:2015qwp,Shannon:2015ect,Arzoumanian:2015liz,Qin:2018yhy}, may yield crucial information about the physics of the early universe much before the hot big bang. However, the search of SGWB in the data often relies on the templates of the GW spectrum one is assuming \cite{Kuroyanagi:2018csn,Caprini:2019pxz}. Thus, it is mandatory to explore the vast range of possible GW sources in the early universe. In this respect, an interesting candidate is the so-called induced GWs generated from primordial density fluctuations \cite{Ananda:2006af,Baumann:2007zm}, which is an essential counterpart to the PBH scenario \cite{Sasaki:2018dmp,Sato-Polito:2019hws} and a probe of the primordial spectrum on scales smaller than those probed by the CMB \cite{Assadullahi:2009jc,Bugaev:2010bb,Inomata:2018epa}.

Recent studies also suggested that the induced GWs could be a probe of the thermal history of the universe \cite{Cai:2019cdl,Hajkarim:2019nbx,Domenech:2019quo}. In particular, the study of the infrared tail of the induced GWs is important for future SGWB searches with a power-law template \cite{Cai:2019cdl}. Thus, in this paper we studied concrete examples of the generation of induced GWs by a primordial density spectrum peaked at a scale $k_*$ in cosmologies with a constant deceleration, or in other words, with an equation of state $w>-1/3$. To do that, we considered a canonical scalar field in an exponential potential, the so-called power-law model \cite{Lucchin:1984yf}. The main reason for this choice is that the propagation speed of perturbations for a canonical scalar field is $c_s^2=1$ independent of $w$, while for a adiabatic perfect fluid it is given by $c_s^2=w$ and becomes negative for $w<0$. 

In Secs.\ref{sec:inducedgws} and \ref{sec:radiationdomination} we have derived analytical formulas for the induced GW spectrum for a general $w>-1/3$. The spectrum is given by Eqs.~\eqref{eq:OMsuperh} and \eqref{eq:phrd}, complemented by Eqs.~\eqref{eq:farsuperh} and \eqref{eq:deepsubh}.
These are the main results of this paper. Furthermore, in Secs.~\ref{subsec:delta} and \ref{subsec:finitewidth} we have studied the GW spectrum generated by a delta Dirac and a finite width peak in the primordial density power spectrum. In particular, we have focused on the infrared tail of the GW spectrum, that is for scales far below the characteristic scale $k\ll k_*$. We obtained that the infrared side of the spectrum is given by
\begin{align}\label{eq:}
\Omega_{\rm GW}(k\ll k_*)\sim\left\{
\begin{aligned}
&k^{3} & (k\lesssim k_{\rm rh})\\
&k^{3-2|\beta|} &(k\gtrsim k_{\rm rh})
\end{aligned}
\right.\,,
\quad {\rm where}\quad \beta=\frac{1-3w}{1+3w}\,,
\end{align}
and in the case of a very sharp peak (or Dirac delta) in the scalar spectrum the GW spectrum has to be multiplied by an additional $k^{-1}$. It should be noted that for $\beta<0$ the scale of the knee seems to be $k\sim k_{\rm rh}/(1+\beta)$ rather than $k_{\rm rh}$ since $k_{\rm rh}/(1+\beta)>k_{\rm rh}$. However, we need numerical calculations to confirm the position of the knee since for $\beta<0$ the superhorizon approximation breaks down faster the negative the beta as one approaches to $k\sim k_{\rm rh}/(1+\beta)$.

The parameter $\beta$ quantifies how much is the background expansion deviates from the radiation dominated universe with $\beta=0$ ($w=1/3$) during the scalar field domination. Note that this factor $\beta$ generally appears when comparing the background expansion with general $w$ to that of radiation domination. Interestingly, we obtained that the infrared spectral index for $k>k_{\rm rh}$ is $3-2|\beta|$ rather than the expected $3-2\beta$ from the relative evolution of the background expansion with respect to the energy density of GWs, which decays as radiation. This difference is due to a superhorizon growth of tensor modes for $\beta<0$ ($w>1/3$) even after the scalar fluctuations have entered the horizon. Such superhorizon growth comes from the fact that the source term for tensor modes, which is a scalar field fluctuation squared and decays as radiation, grows for $\beta<0$. We provided analytical approximations for the infrared GW spectrum in the Dirac delta case in Eqs.~\eqref{eq:OMbn} ($\beta<0$) and \eqref{eq:OMbp} ($\beta>0$). The detailed shape of the spectrum is illustrated in Figs.~\ref{fig:deltasmatching}, \ref{fig:deltasbn} and \ref{fig:deltasbp}. We have also argued that only when $\beta\in\mathbb{Z}$ the GW spectral index in the infrared tail has a logarithmic correction, which includes the radiation domination case ($\beta=0$) discussed in Ref.~\cite{Yuan:2019wwo}. Otherwise, for $\beta\notin\mathbb{Z}$ the logarithmic correction is absent.

In Sec.~\ref{subsec:finitewidth} we studied in more detail the effects of a finite peak width, which constitutes a more realistic scenario than the Dirac delta case. We found that the finiteness of the width introduces a new scale where the GW spectrum slope changes one additional power of $k$, yielding richer structure. To study this case analytically, we considered a log-normal peak with dimensionless width $\sigma\ll1$ \eqref{eq:lognormalpeak}. We obtained that the GW spectrum given by Eq.~\eqref{eq:omegafinitewidth}, presents two possibilities. First, if $\sigma<k_{\rm rh}/k_*$ the change is for scales with $k<k_{\rm rh}$ and the spectrum transitions from $k^{3}$ to $k^{2}$ at $k\sim\sigma\,k_*$. Second and most interesting, if $\sigma>k_{\rm rh}/k_*$ the slope of the infrared tail changes for scales with $k>k_{\rm rh}$ and goes from $k^{3-2|\beta|}$ to $k^{2-2|\beta|}$ at $k\sim\sigma\,k_*$. This is illustrated by Eq.~\eqref{eq:summaryfinitewidth} and Fig.~\ref{fig:finitewidth}.

We noted that the induced SGWB presents a degeneracy in the infrared slope for $|\beta|<1$. However, this degeneracy would be broken by a detailed analysis of the shape of the spectrum around the scales $k\sim k_{\rm rh}/(1+\beta)$ and $k\sim k_{\rm rh}$. For instance, the position of the knee for $\beta>0$ is near $k\sim k_{\rm rh}$ and for $\beta<0$ it seems to be close to $k\sim k_{\rm rh}/(1+\beta)$. This is shown in Fig.~\ref{fig:deltasmatching}. Also, we expect that the derived GWs spectrum would present small oscillations on top of our estimate, imprinting the last few oscillations of scales that entered the horizon right before (re)heating. We see such oscillations for $\beta>0$ in Fig.~\ref{fig:deltasmatching}. For $\beta<0$ numerical calculations are needed since our approximation breaks down faster towards $k\sim k_{\rm rh}/(1+\beta)$ the negative the $\beta$.
This means that the induced GWs power spectrum can be used to test the thermal history of the universe between inflation and the hot big bang.

Before we conclude our work, it is important to analyze possible degeneracies of the infrared region of the induced GWs power spectrum with already known sources of cosmological gravitational waves. We present three examples of degeneracy of power spectra around a characteristic scale $k_o$. First, we see that for $\beta=2$ ($w=-1/9$) the induced GW spectrum of a finite width peak resembles the GW spectrum generated by first order phase transitions (or domain walls), which goes as $k^{2.8}$ (or $k^3$) for $k<k_o$ and $k^{-1}$ for $k>k_o$ (see Ref.~\cite{Kuroyanagi:2018csn} and references therein). Second, we find a similar shape of the induced GW spectrum for $3/2>\beta>0$ to GWs from the Pre-Big-Bang model where the spectrum roughly goes as $k^3$ for $k<k_o$ and $k^{3-2\mu}$ for $k>k_o$ where $3/2>\mu>0$ \cite{Buonanno:1996xc,Kuroyanagi:2017kfx} (although see Ref.~\cite{Gasperini:2016gre} for an updated spectrum). Lastly, we see that the induced GW spectrum generated by a Dirac delta with $3>\beta>0$ is degenerated with that of short-lived global cosmic strings \cite{Kamada:2015iga}. Such short-lived cosmic strings generate a GW spectrum proportional to $k^2$ for $k<k_o$ and $k^\gamma$ for $k>k_o$ where $\gamma=\frac{2n-16}{n-2}$ and $2n-2\geq 6$ is the power of the higher dimensional operators in the potential for the scalar field. However, even in all these cases a closer inspection of the shape of the spectrum near the peak will break the degeneracy due to the oscillations in the induced SGWB. We also find that the induced GW spectrum has the following distinct signatures: $(i)$ it presents oscillations around the scale of reheating, $(ii)$ the infrared tilt for $k>k_{\rm rh}$ can never be bigger than $3$ but $(iii)$ it can be red and infinitely steep.

\section*{Acknowledgments}
G.D. would like to thank T.~Vargas for his hospitality during the visit to the national university of San Marcos while this work was being done and A.~D.~Rojas, D.~Rojas and G.~Neyra for the local support. G.D. was partially supported by the DFG Collaborative Research center SFB 1225 (ISOQUANT) and the European Union’s Horizon 2020 research and innovation programme (InvisiblesPlus) under the Marie Sk{\l}odowska-Curie grant agreement No. 690575. 
The work of S.P. is supported in part by JSPS Grant-in-Aid for Early-Career Scientists No. 20K14461. 
The work of M.S. is supported in part by the JSPS KAKENHI Nos. 19H01895 and 20H04727. S.P. and M.S. are supported in part by the World Premier International Research Center Initiative (WPI Initiative), MEXT, Japan. Calculations of cosmological perturbation theory at second order were checked using the \texttt{xPand (xAct) Mathematica} package.

\appendix

\section{Calculations of perturbations\label{app:formulasperturbations}}
Here we present the necessary equations and formulas to compute the induced spectrum in the Poisson and flat gauges.
\subsection{Poisson gauge\label{app:poissongauge}}
At first order we have that
\begin{align}
\Phi=-\Psi
\end{align}
and that
\begin{align}
\Phi''+\left(2\epsilon-\eta\right){\cal H}\Phi'-\eta{\cal H}^2\Phi+k^2\Phi=0\,.
\end{align}
In the power-law case, this equation is further reduced to
\begin{align}
\Phi''+2\epsilon{\cal H}\Phi'+k^2\Phi=0\,.
\end{align}
\subsection{Flat gauge \label{app:flatgauge}}
The energy momentum tensor of the scalar field is given by
\begin{align}
T_{\mu\nu}=\partial_\mu\phi\partial_\nu\phi-g_{\mu\nu}\left(\frac{1}{2}\partial_\mu\phi\partial^\mu\phi+V(\phi)\right)\,.
\end{align}
The FLRW metric in the flat gauge reads
\begin{align}\label{eq:flat}
ds^2&=a^2(\tau)\left[-(1+2\alpha)d\tau^2+2\beta_idx^idt+\left(\delta_{ij}+2h_{ij}\right)dx^idx^j\right]\,.
\end{align}
We then find that at first order the scalar field fluctuations $\phi\to\phi+\delta\phi$ obey
\begin{align}
\delta\phi''+2{\cal H}\delta\phi'-\Delta\delta\phi+\frac{\eta}{2}{\cal H}^2\delta\phi^2\left(\epsilon-3-\frac{\eta}{2}-\frac{\eta'}{{\cal H}\eta}\right)=0\,,
\end{align}
where we have used that
\begin{align}
\alpha=\frac{\phi'}{2{\cal H}}\delta\phi \quad {\rm and}\quad \Delta\beta=-\frac{\phi'}{2{\cal H}}\delta\phi'-\frac{a^2}{2{\cal H}}\delta\phi\left[V_\phi+\frac{V\phi'}{{\cal H}}\right]\,.
\end{align}
Expanding at second order in perturbation theory one finds that
\begin{align}
^{(2)}T_{ij}^{TT}=\partial_i\delta\phi\partial_j\delta\phi
\end{align}
and
\begin{align}
^{(2)}G_{ij}^{TT}=-\partial_i\alpha\partial_j\alpha-4{\cal H}\partial_i\alpha\partial_j\beta-\partial_i\alpha'\partial_j\beta-2\partial_i\alpha\partial_j\beta'+\partial_k\left(\partial^k\beta\partial_i\partial_j\beta\right)\,,
\end{align}
where the superindex $TT$ refers to transverse-traceless component.

If we specialize to the power-law model, we have that
\begin{align}
\frac{\phi'}{2{\cal H}}=\frac{\lambda}{2}\quad{,}\quad \epsilon=\frac{2+\beta}{1+\beta}\quad{,}\quad\eta=0 \quad {\rm and} \quad \Delta\beta=-\frac{\phi'}{2{\cal H}}\delta\phi'\,.
\end{align}

\section{Matching to radiation domination\label{app:matching}}
In this appendix we present the formulas used to match our solutions during the power-law expansion to the later radiation domination. We assume that there is a sudden transition between the scalar field domination and radiation domination. First of all, we have that scale factor during radiation domination is given by
\begin{align}
a_{RD}(\tau)=a_0\left(\frac{\tau_{\rm rh}}{\tau_0}\right)^{\beta}\left(\frac{(1+\beta)\tau-\beta\tau_{\rm rh}}{\tau_0}\right)\,,
\end{align}
and yields that the conformal hubble parameter reads
\begin{align}
{\cal H}=\frac{1+\beta}{(1+\beta)\tau-\beta\tau_{\rm rh}}\,.
\end{align}

The general solution to the linear tensor modes, namely
\begin{align}
h''+2{\cal H}h'+k^2h=0\,,
\end{align}
is given by
\begin{align}\label{eq:hrd}
h_{RD}(k\tilde \tau)=\frac{C_{1,RD}\sin(k\tilde \tau)+C_{2,RD}\cos(k\tilde \tau)}{k\tilde \tau}\,,
\quad{\rm where}\quad
\tilde\tau\equiv\tau-\frac{\beta}{1+\beta}\tau_{\rm rh}\,.
\end{align}
With this solution we can match both superhorizon and subhorizon modes with the ones during scalar domination. Also, at this point we emphasize the difference between our approximation in Sec.~\ref{sec:inducedgws}. For the solution during radiation domination we see that the superhorizon regime refers to scales which $k\tilde\tau=k/{\cal H}\ll1$, in contrast to the scalar dominated regime where it was more useful to use $k\tau\ll1$. The same logic applies to subhorizon scales.

On one hand, we have that during the scalar field domination the induced tensor modes on superhorizon scales are given by
\begin{align}
h^{\rm induced}_\beta(k\tau\ll k\tau_{\rm rh})=\tilde C_{1,\beta}+\tilde C_{2,\beta}(k\tau)^{-2\beta}\,.
\end{align}
If we focus on the far superhorizon regime in which $k\ll{\cal H}$ we have that Eq.~\eqref{eq:hrd} reduced to
\begin{align}
h_{RD}(k\tau\ll k\tau_{\rm rh})=C_{1,RD}+\frac{C_{2,RD}}{k\tau-\frac{\beta}{1+\beta}k\tau_{\rm rh}}\,.
\end{align}
Requiring that the amplitude and its first derivative are continue we arrive at the matching condition for superhorizon modes, that is
\begin{align}
C_{1,RD}=\tilde C_{1,\beta}+\frac{1-\beta}{1+\beta} \tilde C_{2,\beta}(k\tau_{\rm rh})^{-2\beta}
\end{align}
and
\begin{align}
C_{2,RD}=\frac{2\beta}{\left(1+\beta\right)^2} \tilde C_{2,\beta}(k\tau_{\rm rh})^{1-2\beta}\,.
\end{align}
However, we have seen that there may be situations where $k\tau\ll1$ does not correspond exactly to $k\ll{\cal H}$. In this case, we have that for the general solution \eqref{eq:hrd} the matching yields
\begin{align}
C_{1,RD}=\frac{(k\tau_{\rm rh})^{-2\beta}}{1+\beta}&\Bigg(k\tau_{\rm rh}\left(\tilde C_{2,\beta}+\tilde C_{1,\beta}(k\tau_{\rm rh})^{2\beta}\right)\cos\left[\frac{k\tau_{\rm rh}}{1+\beta}\right]\nonumber\\&-\left((1-\beta)\tilde  C_{2,\beta}+(1+\beta)\tilde  C_{1,\beta}(k\tau_{\rm rh})^{2\beta}\right)\sin\left[\frac{k\tau_{\rm rh}}{1+\beta}\right]\Bigg)
\end{align}
and
\begin{align}
C_{2,RD}=\frac{(k\tau_{\rm rh})^{-2\beta}}{1+\beta}&\Bigg(k\tau_{\rm rh}\left(\tilde  C_{2,\beta}+\tilde C_{1,\beta}(k\tau_{\rm rh})^{2\beta}\right)\sin\left[\frac{k\tau_{\rm rh}}{1+\beta}\right]\nonumber\\&+\left((1-\beta)\tilde C_{2,\beta}+(1+\beta)\tilde C_{1,\beta}(k\tau_{\rm rh})^{2\beta}\right)\cos\left[\frac{k\tau_{\rm rh}}{1+\beta}\right]\Bigg)\,.
\end{align}

On the other hand, we have that subhorizon modes during the scalar field domination go as
\begin{align}
h^{\rm induced}_\beta(k\tau\gg k\tau_{\rm rh})=\sqrt{\frac{2}{\pi}}(k\tau)^{-(\beta+1)}\left(C_{1,\beta}\sin\left[k\tau-\beta\pi/2\right]-C_{2,\beta}\cos\left[k\tau-\beta\pi/2\right]\right)\,.
\end{align}
Matching the tensor modes and their derivative with Eq.~\eqref{eq:hrd} at reheating we find that
\begin{align}
C_{1,RD}=-\sqrt{\frac{2}{\pi}}\frac{(k\tau_{\rm rh})^{-\beta}}{1+\beta}\left(C_{1,\beta}\sin\left[\frac{\beta\pi}{2}-\frac{\beta k\tau_{\rm rh}}{1+\beta}\right]+C_{2,\beta}\cos\left[\frac{\beta\pi}{2}-\frac{\beta k\tau_{\rm rh}}{1+\beta}\right]\right)
\end{align}
and
\begin{align}
C_{2,RD}=\sqrt{\frac{2}{\pi}}\frac{(k\tau_{\rm rh})^{-\beta}}{1+\beta}\left(C_{1,\beta}\cos\left[\frac{\beta\pi}{2}-\frac{\beta k\tau_{\rm rh}}{1+\beta}\right]-C_{2,\beta}\sin\left[\frac{\beta\pi}{2}-\frac{\beta k\tau_{\rm rh}}{1+\beta}\right]\right)\,.
\end{align}
In this case, we have a simplification for the sum of the squares, which yields
\begin{align}
C_{1,RD}^2+C_{2,RD}^2=\frac{2}{\pi}\frac{(k\tau_{\rm rh})^{-2\beta}}{\left(1+\beta\right)^2}\left(C_{1,\beta}^2+C_{2,\beta}^2\right)\,.
\end{align}

Notice that for $\beta\gg1$ the Bessel functions decay slower in conformal time and it will be useful to have the general matching with the general form
\begin{align}
h^{\rm induced}_\beta(k\tau\gg k\tau_{\rm rh})=(k\tau)^{-(\beta+1/2)}\left(C_{1,\beta}J_{\beta+1/2}(k\tau)+C_{2,\beta}Y_{\beta+1/2}(k\tau)\right)\,.
\end{align}
In this case we find that
\begin{align}
C_{1,RD}=\frac{(k\tau_{\rm rh})^{-\beta-1/2}}{1+\beta}&\Bigg(k\tau_{\rm rh}\left(C_{1,\beta} J_{\beta+1/2}(k\tau_{\rm rh})+C_{2,\beta}Y_{\beta+1/2}(k\tau_{\rm rh})\right)\cos\left[\frac{k\tau_{\rm rh}}{1+\beta}\right]\nonumber\\&
-\bigg((1+\beta)\left(C_{1,\beta}J_{\beta+1/2}(k\tau_{\rm rh})+C_{2,\beta}Y_{\beta+1/2}(k\tau_{\rm rh})\right)\nonumber\\&-k\tau_{\rm rh}\left(C_{1,\beta} J_{\beta+3/2}(k\tau_{\rm rh})+C_{2,\beta}Y_{\beta+3/2}(k\tau_{\rm rh})\right)\bigg)\sin\left[\frac{k\tau_{\rm rh}}{1+\beta}\right]\Bigg)
\end{align}
and
\begin{align}
C_{2,RD}=\frac{(k\tau_{\rm rh})^{-\beta-1/2}}{1+\beta}&\Bigg(k\tau_{\rm rh}\left(C_{1,\beta} J_{\beta+1/2}(k\tau_{\rm rh})+C_{2,\beta}Y_{\beta+1/2}(k\tau_{\rm rh})\right)\sin\left[\frac{k\tau_{\rm rh}}{1+\beta}\right]\nonumber\\&
+\bigg((1+\beta)\left(C_{1,\beta}J_{\beta+1/2}(k\tau_{\rm rh})+C_{2,\beta}Y_{\beta+1/2}(k\tau_{\rm rh})\right)\nonumber\\&-k\tau_{\rm rh}\left(C_{1,\beta} J_{\beta+3/2}(k\tau_{\rm rh})+C_{2,\beta}Y_{\beta+3/2}(k\tau_{\rm rh})\right)\bigg)\cos\left[\frac{k\tau_{\rm rh}}{1+\beta}\right]\Bigg)\,.
\end{align}

After we have matched the tensor modes, we have to follow them until they are deep inside the horizon during radiation domination, that is $k\tau\gg k\tau_{\rm rh}$. Then we find that, as usual,
\begin{align}\label{eq:hrd2}
h_{RD}(k\tau\gg k\tau_{\rm rh})\approx\frac{C_{1,RD}\sin(k\tau)+C_{2,RD}\cos(k\tau)}{k\tau}\,.
\end{align}

\section{Bessel functions\label{app:bessel}}

We write below useful formulas related to the Bessel functions. First, the asymptotic expansion for small argument is given by
\begin{align}
{J_\nu(x\ll1)}\approx x^\nu\frac{2^{-\nu}}{\Gamma[1+\nu]}+O(x^{\nu+1})
\quad,\quad
{Y_\nu(x\ll1)}\approx-\frac{2^{\nu}}{\pi}{\Gamma[\nu]}x^{-\nu}+O(x^{-\nu+1})\,.
\end{align}
If $\nu\in \mathbb{Z}$ we have that
\begin{align}
{Y_\nu(x\ll1)}\approx-\frac{2^{\nu}}{\pi}{\Gamma[\nu]}x^{-\nu}+x^\nu 2^{1-\nu}\frac{\gamma_E+\ln(x/2)-H_\nu/2}{\pi\Gamma[1+\nu]} +O(x^{-\nu+1})\,,
\end{align}
where
\begin{align}
H_\nu=\sum_{n=1}^{\nu}\frac{1}{n}\,.
\end{align}
This formula will only be relevant when $\nu=0$ as for other values of $\nu>0$ the second term will be suppressed. 

For large arguments we have that the Bessel functions oscillate periodically as
\begin{align}
{J_\nu(x\gg1)}\approx\sqrt{\frac{2}{\pi x}}\cos\left(x-\frac{\nu\pi}{2}-\frac{\pi}{4}\right)+O(x^{-1})
\end{align}
and
\begin{align}
{Y_\nu(x\gg1)}\approx\sqrt{\frac{2}{\pi x}}\sin\left(x-\frac{\nu\pi}{2}-\frac{\pi}{4}\right)+O(x^{-1})\,.
\end{align}

Other useful relations between derivative and Bessel functions of similar order are given by
\begin{align}
\partial_x{J}_{\nu}\left(x\right)={J}_{\nu-1}\left(x\right)-(\nu/x)%
{J}_{\nu}\left(x\right)\,,
\end{align}
and
\begin{align}
{J}_{\nu-1}\left(x\right)+{J}_{\nu+1}\left(x\right)=(2\nu/x)%
{J}_{\nu}\left(x\right)\,.
\end{align}

\section{Formulas for \texorpdfstring{$\beta=-1/2$}{case}. \label{app:formulasparticularbeta}}

In this appendix we present the formulas for the $\beta=-1/2$ ($w=1$) case. We have to treat this value of $\beta$ separately due to the special behavior of the Bessel functions of order $0$, specially the Bessel function of the second kind diverges logarithmically for small argument stead of a power-law. We list below some useful formulas.

The averaged kernel squared on sub-horizon scales goes as
\begin{align}\label{eq:kernellimit12}
\overline{I^2(v\gg1,\beta=-1/2,x\gg1)}\approx \frac{9}{\pi x v^2}\,.
\end{align}
On super horizon scales, the Green function involves a logarithm and therefore, we need to treat the integral separately. Thus we will focus on the kernel directly which is given by
\begin{align}\label{eq:kernel3}
I(u,v,\beta,x)=\frac{3\pi}{4}{\cal I}(u,v,\beta=-1/2,x)\,,
\end{align}
where
\begin{align}\label{eq:int3}
{\cal I}
(v,v,\beta=-1/2,x)=\frac{2}{\pi v^2}\int_0^{vx} d\hat x \,\ln\left(\frac{vx}{\hat x}\right)\hat x
\left[J_{0}(\hat x)J_{0}(\hat x)+3J_{2}(\hat x)J_{2}(\hat x)\right]\,.
\end{align}
Then, we find that on superhorizon scales the kernel is approximately given by
\begin{align}
I(v,\beta=-1/2,x\ll1,vx\gg1)\approx C_1(k)+C_2(k) x
\end{align}
where
\begin{align}
C_1(k)=-\frac{9}{2}v^{-2}\quad {\rm and}\quad C_2(k)=\frac{6}{\pi}v^{-1} \,.
\end{align}
Lastly, for a Dirac delta spectrum we find that on subhorizon scales the GW spectrum is given by
\begin{align}
\Omega_{\rm GW,c}(\beta=-1/2,k_*\gg k \gg k_{\rm rh})=\frac{3{\cal A}^2_{\cal R}}{2\pi^3}\left(\frac{k_*}{k_{\rm rh}}\right)\left(\frac{k}{k_{*}}\right)
\end{align}
and on superhorizon scales by
\begin{align}
\Omega_{\rm GW,c}(\beta=-1/2,k_*\gg k_{\rm rh}\gg k)=\frac{6{\cal A}^2_{\cal R}}{\pi^2}\left(\frac{k_*}{k_{\rm rh}}\right)^{2}\left(\frac{k}{k_{*}}\right)^2\,.
\end{align}

\section{Correction terms for the sub-horizon approximation \label{app:subhorizon}}

In this appendix we show that the subhorizon approximation used in Sec.~\ref{subsec:subh1} is indeed value for any value of $\beta$. First, we divide the integral into two parts
\begin{align}
{\cal I}^x_{J,Y}(u,v,\beta,x)={\cal I}^\infty_{J,Y}(u,v,\beta)+\Delta{\cal I}^x_{J,Y}(u,v,\beta,x)
\end{align}
where
\begin{align}\label{eq:int1}
{\cal I}^\infty_{J,Y}
(u,v,\beta)\equiv\int_0^{\infty}& d\tilde x \tilde x^{1/2-\beta}
\left\{	
\begin{aligned}
	J_{\beta+1/2}(\tilde x)\\
	Y_{\beta+1/2}(\tilde x)
\end{aligned}
\right\}\nonumber\\&\times
\left[J_{\beta+1/2}(u\tilde x)J_{\beta+1/2}(v\tilde x)+\frac{2+\beta}{1+\beta}J_{\beta+5/2}(u\tilde x)J_{\beta+5/2}(v\tilde x)\right]\,,
\end{align}
and
\begin{align}\label{eq:int1}
\Delta{\cal I}_{J,Y}
(u,v,\beta,x)\equiv\int_\infty^{x} &d\tilde x \tilde x^{1/2-\beta}
\left\{	
\begin{aligned}
	J_{\beta+1/2}(\tilde x)\\
	Y_{\beta+1/2}(\tilde x)
\end{aligned}
\right\}\nonumber\\&\times
\left[J_{\beta+1/2}(u\tilde x)J_{\beta+1/2}(v\tilde x)+\frac{2+\beta}{1+\beta}J_{\beta+5/2}(u\tilde x)J_{\beta+5/2}(v\tilde x)\right]\,.
\end{align}

We can then evaluate the error we are making by computing $\Delta{\cal I}_{J,Y}$ in the limit of large argument for $v\sim u$. In this way, we find that
\begin{align}\label{eq:int1}
\Delta{\cal I}_{J,Y}
(u,v,\beta,x\gg1)\approx&-\frac{3+2\beta}{1+\beta}\left(\frac{2}{\pi}\right)^{3/2}(uv)^{-1/2}\nonumber\\&\times\int_\infty^x d\tilde x \tilde x^{-1-\beta}
\left\{	
\begin{aligned}
\sin\left(\varphi-\tilde x\right)\\
\cos\left(\varphi-\tilde x\right)
\end{aligned}
\right\}
\sin\left(\varphi-u\tilde x\right)\sin\left(\varphi-v\tilde x\right)
\end{align}
where 
\begin{align}
\varphi\equiv \frac{\beta\pi}{2}\,.
\end{align}

After integration we find that
\begin{align}\label{eq:int1}
\Delta{\cal I}_{J}
(v,v,\beta,x\gg1)\approx&-\frac{1}{4}\frac{3+2\beta}{1+\beta}\left(\frac{2}{\pi}\right)^{3/2}(uv)^{-1/2}x^{-1-\beta}\nonumber\\&\times
\Bigg(\frac{\cos\left[\varphi-(1-u+v)x\right]}{1-u+v}+\frac{\cos\left[\varphi-(1+u-v)x\right]}{1+u-v}\nonumber\\&+\frac{\cos\left[\varphi+(1-u-v)x\right]}{1-u-v}+\frac{\cos\left[3\varphi-(1+u+v)x\right]}{1-u+v}\Bigg)
\end{align}
and
\begin{align}\label{eq:int1}
\Delta{\cal I}_{Y}
(u,v,\beta,x\gg1)\approx&-\frac{1}{4}\frac{3+2\beta}{1+\beta}\left(\frac{2}{\pi}\right)^{3/2}(uv)^{-1/2}x^{-1-\beta}
\nonumber\\&\times\Bigg(\frac{\sin\left[\varphi-(1-u+v)x\right]}{1-u+v}+\frac{\sin\left[\varphi-(1+u-v)x\right]}{1+u-v}\nonumber\\&+\frac{\sin\left[\varphi+(1-u-v)x\right]}{1-u-v}+\frac{\sin\left[3\varphi-(1+u+v)x\right]}{1-u+v}\Bigg)
\end{align}
We see that in general
\begin{align}
\Delta{\cal I}_{J,Y}(v,v,\beta,x\gg1)\propto v^{-1}x^{-1-\beta}\ll {\cal I}^{\infty}_{J,Y}(v,v,w)\propto v^{-1},v^{2\beta-1}\,.
\end{align}

\section{Analytic integrals with three Bessel functions \label{App:integralbessel}}

We review here the results of Ref.~\cite{threebesselI}. They find that for $|a-b|<c<a+b$ and $\beta>-1$
\begin{align}
\int_0^{\infty} d\tilde x \,\tilde x^{1/2-\beta}
\left\{	
\begin{aligned}
	J_{\beta+1/2}(c\tilde x)\\
	Y_{\beta+1/2}(c\tilde x)
\end{aligned}
\right\}
J_{\nu+1/2}(a\tilde x)J_{\nu+1/2}(b\tilde x)=\frac{1}{\pi}\sqrt{\frac{2}{\pi}}\frac{(ab)^{\beta-1/2}}{c^{\beta+1/2}}\left(\sin\varphi\right)^{\beta}\left\{	
\begin{aligned}
	\frac{\pi}{2}\mathsf{P}^{-\beta}_{\nu}(\cos\varphi)\\
	-\mathsf{Q}^{-\beta}_{\nu}(\cos\varphi)
\end{aligned}
\right\}
\end{align}
where
\begin{align}
16\Delta^2\equiv\left(c^2-(a-b)^2\right)\left((a+b)^2-c^2\right)\quad,\quad
\cos\varphi=\frac{a^2+b^2-c^2}{2ab}\quad,\quad\sin\varphi=\frac{2\Delta}{ab}\,.
\end{align}

We can use these formulas identifying
\begin{align*}
c=1\quad,\quad a=u\quad,\quad b=v\,.
\end{align*}
In that case the range $|u-v|<1<u+v$ covers all range of interest. 

\section{Legendre functions on the cut \label{App:legendrefunctions}}

The Legendre functions on the cut are defined for $|y|<1$ as
\begin{align}
\mathsf{P}^{\mu}_{\nu}\left(y\right)=\left(\frac{1+y}{1-y}\right)^{\mu/2}%
\mathbf{F}\left(\nu+1,-\nu;1-\mu;\tfrac{1}{2}-\tfrac{1}{2}y\right)\,,
\end{align}
\begin{align}
\mathsf{Q}^{\mu}_{\nu}\left(y\right)=\frac{\pi}{2\sin\left(\mu\pi\right)}&\Bigg\{%
\cos\left(\mu\pi\right)\left(\frac{1+y}{1-y}\right)^{\mu/2}\mathbf{F}\left(%
\nu+1,-\nu;1-\mu;\tfrac{1}{2}-\tfrac{1}{2}y\right)\\&-\frac{\Gamma\left(\nu+\mu+1%
\right)}{\Gamma\left(\nu-\mu+1\right)}\left(\frac{1-y}{1+y}\right)^{\mu/2}%
\mathbf{F}\left(\nu+1,-\nu;1+\mu;\tfrac{1}{2}-\tfrac{1}{2}y\right)\Bigg\}\,,
\end{align}
where
\begin{align}
\mathbf{F}\left(a,b;c;y\right)=\frac{1}{\Gamma\left(c\right)}F\left(a,b;c;y\right)
\end{align}
and $F\left(a,b;c;y\right)$ is the Gauss's hypergeometric function.

\subsection{Integer degree and order}
For integer numbers ($m,n>0$) we have that
\begin{align}
\left\{
\begin{aligned}
\mathsf{P}^{m}_{n}(x)\\
\mathsf{Q}^{m}_{n}(x)
\end{aligned}
\right\}=
(-1)^m(1-x^2)^{m/2}\frac{d^m}{dx^m}
\left\{
\begin{aligned}
\mathsf{P}_{n}(x)\\
\mathsf{Q}_{n}(x)
\end{aligned}
\right\}\,,
\end{align}
where
\begin{align}
\mathsf{P}_{n}(x)=\frac{(-1)^n}{2^n n!}\frac{d^n}{dx^n}(1-x^2)^n\,,
\end{align}
and
\begin{align}
\mathsf{Q}_{n}(x)=\frac{1}{2}\mathsf{P}_{n}(x)\ln\left(\frac{1+x}{1-x}\right)-W_{n-1}(x)\,,
\end{align}
with
\begin{align}
W_{n-1}(x)=\sum_{j=1}^n\frac{1}{j}\mathsf{P}_{j-1}(x)\mathsf{P}_{n-j}(x)\,.
\end{align}
Furthermore, we use the relation
\begin{align}
\left\{
\begin{aligned}
\mathsf{P}^{-m}_{n}(x)\\
\mathsf{Q}^{-m}_{n}(x)
\end{aligned}
\right\}=
(-1)^m\frac{\Gamma(n-m+1)}{\Gamma(n+m+1)}
\left\{
\begin{aligned}
\mathsf{P}^m_{n}(x)\\
\mathsf{Q}^m_{n}(x)
\end{aligned}
\right\}\,,
\end{align}
since in the cases under study we always have $\mathsf{P}^{-m}_{n}$ and $\mathsf{Q}^{-m}_{n}$ with $m,n>0$ and $n-m>0$. 

\subsection{Limiting behavior}

We have that for general $\beta$ (except for $\beta=0$) the Legendre function on the cut of the first kind behave as
\begin{align}
\mathsf{P}^{-\beta}_{\beta}(y\sim1)\sim\mathsf{P}^{-\beta}_{\beta+2}(y\sim1)\sim \frac{1}{\Gamma[\beta+1]}\left(\frac{1-y}{2}\right)^{\beta/2}\,.
\end{align}

First, for $\beta>0$ we have that
\begin{align}
\mathsf{Q}^{-\beta}_{\beta}(y\sim1,\beta>0)\sim \frac{\Gamma[2\beta+3]}{2\Gamma[2\beta+1]} \mathsf{Q}^{-\beta}_{\beta+2}(y\sim1,\beta>0)\sim \frac{\Gamma[\beta]}{2\Gamma[2\beta+1]}\left(\frac{1-y}{2}\right)^{-\beta/2}
\end{align}
and for $\beta<0$ (except $\beta=-1/2$)
\begin{align}
\mathsf{Q}^{-\beta}_{\beta}(y\sim1,\beta<0)\sim \mathsf{Q}^{-\beta}_{\beta+2}(y\sim1,\beta<0)\sim -\frac{1}{2}\cos(\beta\pi)\Gamma[-\beta]\left(\frac{1-y}{2}\right)^{\beta/2}\,.
\end{align}

\section{Kernels for particular cases \label{App:kernelsparticular}}

We present in this appendix the expression of Legendre functions on the cut for $\beta$ integers or half-integers in the range of interest.

\paragraph{$\mathbf{\beta=-1/2}$ ($\mathbf{w=1}$):}
\begin{align}
\mathsf{P}^{1/2}_{-1/2}(y)&=\sqrt{\frac{2}{\pi}} (1 - y^2)^{-1/4}\quad,\quad \mathsf{P}^{1/2}_{3/2}(y)=\sqrt{\frac{2}{\pi}} (1 - y^2)^{-1/4}\left(-1+2y^2\right)\,,\\
\mathsf{Q}^{1/2}_{-1/2}(y)&=0\quad ,\quad \mathsf{Q}^{1/2}_{3/2}(y)=-\sqrt{2\pi}y(1 - y^2)^{1/4}\,.
\end{align}

\paragraph{$\mathbf{\beta=0}$  ($\mathbf{w=1/3}$):}

\begin{align}
\mathsf{P}^{0}_{0}(y)&=1\quad,\quad \mathsf{P}^{0}_{2}(y)=\frac{1}{2}(-1+3y^2)\,,\\
\mathsf{Q}^{0}_{0}(y)&=\frac{1}{2}\ln\left[\frac{1+y}{1-y}\right]\quad ,\quad \mathsf{Q}^{0}_{2}(y)=\frac{1}{4}\left(-6y+(-1+3y^2)\ln\left[\frac{1+y}{1-y}\right]\right)\,.
\end{align}

\paragraph{$\mathbf{\beta=1/2}$ ($\mathbf{w=1/9}$):}

\begin{align}
\mathsf{P}^{-1/2}_{1/2}(y)&=\sqrt{\frac{2}{\pi}} (1 - y^2)^{1/4}\quad,\quad \mathsf{P}^{-1/2}_{5/2}(y)=\frac{1}{3}\sqrt{\frac{2}{\pi}} (1 - y^2)^{1/4}(-1+4y^2)\,,\\
\mathsf{Q}^{-1/2}_{1/2}(y)&=\sqrt{\frac{\pi}{2}}\frac{y}{(1 - y^2)^{1/4}}\quad ,\quad \mathsf{Q}^{-1/2}_{5/2}(y)=\frac{1}{3}\sqrt{\frac{\pi}{2}}\frac{y}{(1 - y^2)^{1/4}}(-3+4y^2)\,.
\end{align}

\paragraph{$\mathbf{\beta=1}$ ($\mathbf{w=0}$):}

\begin{align}
\mathsf{P}^{-1}_{1}(y)&=\frac{1}{2}\sqrt{1 - y^2}\quad,\quad \mathsf{P}^{-1}_{3}(y)=\frac{1}{8}\sqrt{1 - y^2}\left(-1+5y^2\right)\,,\\
\mathsf{Q}^{-1}_{1}(y)&=\frac{2y+(1-y^2)\ln\left[\frac{1+y}{1-y}\right]}{4\sqrt{1-y^2}}\quad ,\quad \mathsf{Q}^{-1}_{3}(y)=\frac{-26y+30y^3-3(1-6y^2+5y^4)\ln\left[\frac{1+y}{1-y}\right]}{48\sqrt{1-y^2}}\,.
\end{align}

\paragraph{$\mathbf{\beta=2/3}$ ($\mathbf{w=-1/15}$):}

\begin{align}
\mathsf{P}^{-3/2}_{3/2}(y)=\frac{1}{3} \sqrt{\frac{2}{\pi}} (1 - y^2)^{3/4}\quad,\quad \mathsf{P}^{-3/2}_{7/2}(y)=\frac{1}{15} \sqrt{\frac{2}{\pi}} (1 - y^2)^{3/4}\left(-1+6y^2\right)\,,\\
\mathsf{Q}^{-3/2}_{3/2}(y)=- \frac{1}{6}\sqrt{\frac{\pi}{2}} \frac{y\left(-3+2y^2\right)}{(1 - y^2)^{3/4}}\quad ,\quad \mathsf{Q}^{-3/2}_{7/2}(y)=- \frac{1}{60}\sqrt{\frac{\pi}{2}} \frac{y\left(15-40y^2+24y^4\right)}{(1 - y^2)^{3/4}}\,.
\end{align}

\paragraph{$\mathbf{\beta=2}$ ($\mathbf{w=-1/9}$):}

\begin{align}
\mathsf{P}^{-2}_{2}(y)&=\frac{1}{8}\left(1-y^2\right)\quad,\quad \mathsf{P}^{-2}_{4}(y)=\frac{1}{48}\left(1-y^2\right)\left(-1+7y^2\right)\,,\\
\mathsf{Q}^{-2}_{2}(y)&=\frac{y \left(5 - 3 y^2\right)}{24 (1 - y^2)} + 
\frac{1}{16} \left(1 - y^2\right) \ln\left[\frac{1+y}{1-y}\right]\,,\\ \mathsf{Q}^{-2}_{4}(y)&=-\frac{y \left(81 - 190 y^2+105y^4\right)}{720 (1 - y^2)} - 
\frac{1}{96} \left(1 - 8y^2+7y^4\right) \ln\left[\frac{1+y}{1-y}\right]\,.
\end{align}

\bibliography{biblio.bib} 

\end{document}